\documentclass[useAMS,usenatbib]{mn2e}
\usepackage{graphicx}
\usepackage{latexsym,amsfonts,amsmath,amssymb}
\usepackage{color}
\usepackage[normalem]{ulem}
\usepackage{hyperref}
\usepackage{subfig}
\usepackage{tabularx}
\usepackage[utf8]{inputenc}
\usepackage{longtable}

\title[CODEX Weak Lensing: Concentration of Galaxy Clusters at z$\sim$0.5]{CODEX Weak Lensing: Concentration of Galaxy Clusters at z$\sim$0.5}
\author[N. Cibirka et al.]{N. Cibirka$^{1,2}$\thanks{E-mail: ncibirka@usp.br (NC)}, E. S. Cypriano$^{1}$, F. Brimioulle$^{3,4,2}$, D. Gruen$^{5,6,20}$, T. Erben$^{7}$\newauthor L. van Waerbeke$^{8}$, L. Miller$^{9}$, A. Finoguenov$^{10,3}$, C. Kirkpatrick$^{10,11}$, J. Patrick Henry$^{12}$\newauthor E. Rykoff$^{5,6}$, E. Rozo$^{13}$, R. Dupke$^{4,14,15,16}$, J-P. Kneib$^{17,18}$, H. Shan$^{17}$,  P. Spinelli$^{19}$\\
$^{1}$Instituto de Astronomia, Geof\'isica e Ci\^encias Atmosf\'ericas, Universidade de S\~ao Paulo, Brasil\\
$^{2}$Universitats-Sternwarte, Fakult\"at f\"ur Physik, Ludwig-Maximilians-Universit\"at M\"unchen, Scheinerstr. 1, D-81679 M\"unchen, Germany\\
$^{3}$Max-Planck-Institute for Extraterrestrial Physics, Giessenbachstra\ss e, 85748 Garching, Germany \\
$^{4}$Observat\'orio Nacional, Rua Gal. Jos\'e Cristino, 20921-400, Rio de Janeiro, Brasil \\
$^{5}$Kavli Institute for Particle Astrophysics \& Cosmology, P. O. Box 2450, Stanford University, Stanford, CA 94305, USA\\
$^{6}$SLAC National Accelerator Laboratory, Menlo Park, CA 94025, USA\\
$^{7}$Argelander-Institut fur Astronomie Auf dem H\"ugel 71, D-53121 Bonn, German\\
$^{8}$Department of Physics and Astronomy, University of British Columbia, 6224 Agricultural road, Vancouver, BC V6T 1Z1, Canada\\
$^{9}$Department of Physics, Oxford University, Keble Road, Oxford OX1 3RH, UK\\
$^{10}$Department of Physics, University of Helsinki, PO Box 64, FI-00014 Helsinki, Finland\\
$^{11}$Helsinki Institute of Physics, Gustaf H\"allstr\"omin kata 2, University of Helsinki, Helsinki, Finland\\
$^{12}$Institute for Astronomy, 2680 Woodlawn Drive, Honolulu, HI 96822, USA\\
$^{13}$Department of Physics, University of Arizona, 1118 E. Fourth St, Tucson, AZ 85721, USA\\
$^{14}$Department of Astronomy, University of Michigan, 311 West Hall
1085 South University Ave. Ann Arbor, MI 48109-1107 USA\\
$^{15}$Department of Physics and Astronomy, University of Alabama, Box 870324, Tuscaloosa, AL 35487, USA\\
$^{16}$Eureka Scientific Inc., 2452 Delmer St. Suite 100, Oakland, CA 94602, USA)\\
$^{17}$Laboratoire d’Astrophysique, Ecole Polytechnique F\`ed\`erale de Lausanne (EPFL), Observatoire de Sauverny, CH-1290 Versoix, Switzerland\\
$^{18}$Aix Marseille Université, CNRS, LAM (Laboratoire d’Astrophysique de Marseille) UMR 7326, 13388, Marseille, France\\
$^{19}$Museu de Astronomia e Ci\^encias Afins (MAST), Rua General Bruce 586, 20921-030 Rio de Janeiro, Brasil\\
$^{20}$Einstein Fellow}

\date{Accepted XXXXXXX. Received 2016 July 05; in original form 2016 July 05}

\pubyear{2016}

\begin{document}
\label{firstpage}
\pagerange{\pageref{firstpage}--\pageref{lastpage}} 
\maketitle

\begin{abstract}

We present a stacked weak lensing analysis of 27 richness selected galaxy clusters at $0.40 \leqslant z \leqslant 0.62$ in the CODEX survey. The fields were observed in 5 bands with the CFHT. We measure the stacked surface mass density profile with a $14\sigma$ significance in the radial range $0.1 < R\ Mpc\ h^{-1} < 2.5$. The profile is well described by the halo model, with the main halo term following an NFW profile and including the off-centring effect. We select the background sample using a conservative colour-magnitude method to reduce the potential systematic errors and contamination by cluster member galaxies. We perform a Bayesian analysis for the stacked profile and constrain the best-fit NFW parameters $M_{200c} = 6.6^{+1.0}_{-0.8} \times 10^{14} h^{-1} M_{\odot}$ and $c_{200c} = 3.7^{+0.7}_{-0.6}$. The off-centring effect was modelled based on previous observational results found for redMaPPer SDSS clusters. Our constraints on $M_{200c}$ and $c_{200c}$ allow us to investigate the consistency with numerical predictions and select a concentration-mass relation to describe the high richness CODEX sample. Comparing our best-fit values for $M_{200c}$ and $c_{200c}$ with other observational surveys at different redshifts, we find no evidence for evolution in the concentration-mass relation, though it could be mitigated by particular selection functions. Similar to previous studies investigating the X-ray luminosity-mass relation, our data suggests a lower evolution than expected from self-similarity.

\end{abstract}


\section{Introduction}
\label{int}

\ \ Dark matter halos, and in particular the halos associated with galaxy clusters, have a prominent position in the $\Lambda$CDM paradigm of formation and evolution of structures in the Universe. As the biggest objects to have reached dynamical equilibrium, galaxy clusters play a central role both in providing cosmological information as well as astrophysical properties. The parameters characterising a given density profile can be related to cosmological constraints, as the characteristic density of a dark matter halo traces the background density of the Universe at the time of collapse \citep{NFW97, Dolag04, Ludlow13}. 
Also through the cluster mass function we can assess cosmological parameters such as $\Omega_m$ and $\sigma_8$ by comparing observations with predictions from simulations of structure formation (e.g. \citealt{White93, Bahcall98, Henry00, Reiprich02, Henry09}). From the astrophysical point of view, clusters are powerful environments to study galaxy evolution, plasma physics and thermodynamics of the intracluster medium and also to test models of gravity \citep{Kravtsov12}. 

N-body simulations based on the $\Lambda$CDM scenario exhibit a universal density profile for dark matter halos. A widely used density profile is the Navarro-Frenk-White profile (hereafter NFW; \citealt{NFW96, NFW97}), described by a scale radius $r_s$ and a characteristic over-density $\delta_c$; this profile can be re-written in terms of the mass of the halo and its concentration parameter $c_{\Delta}$ given by the ratio between a external radius and a scale radius: $c_{\Delta} = r_{\Delta}/r_s$. While the external radius defines a region within which the mean density of the halo is $\Delta$ times the critical density of the universe, the scale radius comes from the radial density profile and for a NFW profile it corresponds to the region where the logarithmic slope changes from $-1$ to $-3$.

From numerical simulations we observe a relation between the concentration parameter and the mass and redshift of the halo. Different groups have found that the concentration decreases with an increase in mass (for a fixed redshift) and redshift (for a fixed mass) \citep{Bullock01, Zhao03, Neto07}. In a hierarchical CDM scenario this trend can be explained as a reflection of the halo assembly epoch - those assembled earlier have a higher concentration following a higher background density. 
This behaviour between the values of mass and concentration of a halo, the concentration-mass relation (hereafter c-M), gives important cosmological information; theories hold the idea that the relation is associated with the mass assembly history of the halo \citep{NFW97, Bullock01, Zhao09,  Giocoli12, Correa15}. As the mass assembly history is correlated with the initial density peak \citep{Dalal08}, some studies have suggested that a relation between concentration and peak height $\nu$ may be more accurate than the c-M relation \citep{Bhattacharya13, Ludlow14, Klypin16} to describe the evolution of halo concentration.

Different simulations do not show perfect agreement on the predictions for the c-M relation (e.g., \citealt{Duffy08, Bhattacharya11, Klypin11, Prada12, Klypin14, Ludlow13, Dutton14, Correa15}, but within the differences between the results, these simulations found that for low redshift the slope of the relation is approximately $-0.1$, considering that we can describe $c \approx A(M/M_{pivot})^{\alpha}$ where $M_{pivot}$ is the median halo mass.

Weak gravitational lensing is a unique tool to constrain cluster masses and to probe the internal structure of dark matter halos: it does not depend on any assumption about the dynamical state of the halo and is sensitive to the total matter content. Studies using lensing analyses found that the NFW profile is a good fit to the  data, especially when fitting a stacked sample of clusters (e.g. \citealt{Mandelbaum06, Umetsu11a, Okabe13, Umetsu14}). The c-M relation for different samples of galaxy clusters has been derived by many of these studies. \citet{Okabe10}, for instance, found a slope of $\alpha \simeq -0.4 \pm 0.19$ from a weak lensing analysis of 19 clusters with $\bar{z} = 0.24$, \citet{SerCov13} inferred $\alpha \simeq -0.83 \pm 0.39$ by analysing 31 weak-lensing systems at high redshift ($\sim 1$). Recently \citet{Groener16} found a slope of $\alpha \simeq -0.38 \pm 0.03$ for a weak lensing sample of 111 systems available in the literature with $\bar{z} = 0.48$. Combining weak and strong lensing analyses \citet{Oguri12} derived a value of $\alpha \simeq -0.59 \pm 0.12 $ for 28 clusters spanning a range of $0.28 < z < 0.68$ while the CLASH collaboration \citep{Clash16} determined a slope of $\alpha \simeq -0.32 \pm 0.18$ for a sample between $0.19 < z < 0.89$. As can be seen these observational studies have found a steeper c-M relation in comparison to the theoretical expectation value of $\alpha = -0.1$. 

Possible explanations for the observed slope of the c-M relation are astrophysical processes such as the baryonic cooling effect \citep{Fedeli12} or supernova and AGN feedback \citep{Duffy10, Mead10}. Other potential sources of deviations between observations and predictions are selection effects \citep{Oguri12, Meneghetti14, Sereno15} or the halo triaxiality that can bias and scatter the projected mass density profile \citep{Meneghetti10, Giocoli12a, Sereno14}. 

However, not all observational studies are in tension with the simulations: \citet{ComNat07} compiled all the concentrations available in the literature at the time - including strong and weak lensing, X-ray and dynamical determinations for clusters up to $z \sim 0.8$ - plus 10 new strong lensing systems and found $\alpha \simeq -0.15 \pm 0.13$; \citet{Mandelbaum08} using data from the Sloan Digital Sky Survey (SDSS) in the range $0.1 < z < 0.3$ derived a slope of $\alpha \simeq -0.13 \pm 0.07$. 

This lack of agreement between different observational studies and numerical simulations shows the need of a better understanding of possible sources of systematic effects in lensing analyses (such as selection effects and shape measurement uncertainties) as well as the inclusion of baryonic physical processes in the simulations. In this paper we will explore the c-M relation on the relatively unexplored window of redshift $z\sim0.5$ with the novel COnstrain Dark Energy with X-ray galaxy clusters (CODEX) sample.

The paper is organised as follows: §\ref{codex} gives an overview of the CODEX survey, §\ref{data} describes the data and catalogues, §\ref{wl} presents the weak lensing analysis detailing the model adopted, the profile fitting, results from the Bayesian analysis and comparison with numerical predictions and other observational studies. We conclude and summarise in §\ref{conclusion}. 

We adopt a concordance $\Lambda$CDM cosmology with $\Omega_m = 0.27$, $\Omega_{\Lambda} = 0.73$ and $H_0 = 100 \ h \ \rm{km}/\rm{s}/\rm{Mpc}$ and consider the overdensity parameter $\Delta$ to be 200 times the critical density of the Universe (200c) throughout the analysis. Masses and concentrations are given for $\Delta = 200c$. Best-fit values are quoted as the median of the posterior distribution and errors refer to the interval containing $68 \%$ of the points, unless otherwise specified.

\section{The CODEX survey}
\label{codex}

\ \ The CODEX survey comprises a sample of galaxy clusters identified as photon overdensities in the ROSAT All Sky Survey (RASS, \citealt{Voges99}), corresponding to a $4\sigma$ photon excess inside a wavelet aperture, and associated with red-sequence galaxies using optical data from the 8th SDSS data release (DR8, \citealt{Aihara11}). The optical counterpart allows to separate galaxy clusters from point sources and assign a photometric redshift as well as an estimate of the richness. A detailed description of the cluster detection and selection are given in Finoguenov et al. (in prep.).

To build the cluster catalogues we have run the redMaPPer algorithm \citep{Rykoff14}, which searches for galaxies corresponding to a cluster red-sequence in the SDSS data around each identified RASS source. The algorithm also provides an estimate for the photometric redshift ($z_{\rm{phot}}$) using galaxy colours and a richness estimator ($\lambda_{\rm{SDSS}}$) corresponding to the sum of the membership probabilities of galaxies for a given identified cluster. The cluster centre is also assigned by redMaPPer and must be located within 3 arcmin from the X-ray position.

The CODEX sample is the largest statistically complete sample of RASS clusters at $z < 0.6$, comprising over 6,600 sources in the SDSS DR8 footprint having a minimal richness of 10. The X-ray limiting flux is $\sim > 2 \times 10^{-13} \rm{ergs}\ \rm{s}^{-1}\ \rm{cm}^{-2}$ with rest-frame X-ray luminosities (L$_X$) in the $[0.1 - 2.4]$ keV energy band computed within R$_{500}$. This leads to a significant probability of detecting clusters with at least 5 photon counts. A total of $\approx 300$ clusters with richness $\lambda_{\rm{SDSS}} \geqslant 60$ were identified in a redshift range of $0.15 < z < 0.6$. The first catalogue of spectroscopically confirmed clusters is published in \citet{Clerc16}. 

The CODEX clusters have a strong overlap with the MAssive Cluster Sample (MACS, \citealt{Ebeling01}) and Planck clusters at $z < 0.4$ and represent a unique sample at $z > 0.4$.



\subsection{CODEX weak lensing sample}

\ \ A subsample of clusters was selected from the initial CODEX sample for weak lensing (WL) follow-up with the Canada-France-Hawaii Telescope (CFHT). The WL follow-up was designed to provide accurate mass measurements to calibrate the mass-observable relations for this survey. The fields were observed between 2012 and 2015 with the wide field optical imaging camera MegaCam \citep{Boulade03}. 

The CFHT follow-up observations can be divided in three subsamples with distinct selection functions: a \textbf{primary} sample of clusters selected by the redMaPPer richness and photometric redshift estimates based on SDSS photometric catalogs, cut at $\lambda_{\rm{SDSS}} \geqslant 60$ and $z_{\rm{ SDSS}} \geqslant 0.4$; a \textbf{secondary} sample with clusters selected only by their ROSAT photon excess and falling inside the CFHT Legacy Survey footprint (CFHTLS, http://www.cfht.hawaii.edu/Science/CFHTLS); a \textbf{tertiary} sample with more complex selection functions including additional CODEX clusters present in the pointed CFHT observations of primary clusters. We note that some clusters belong to both the primary and secondary sample.

Besides the redshift and richness cut the WL subsample selection was also based on technical limitations. Previous experience with the CFHTLS have shown that the presence of a bright star in the vicinity of the target cluster precludes accurate WL studies. So we excluded any CODEX cluster with a 10 (12) mag star within 6’ (2’) of its centre from the WL sample. We also were unable to observe some RA ranges due to conflicts with higher priority CFHT large programs. $\sim 100$ fields survived the above cuts out of which 69 were observed in at least one filter.

The WL subsample studied in this paper consists of clusters from the \textbf{primary} sample, with successful pointed CFHT observations completed in five bands, based only in the observability. The goal is to study the concentration-mass relation of the high richness ($\lambda_{\rm{SDSS}} \geqslant 60$), high redshift (z$ \geqslant 0.4$) CODEX clusters. The photometric redshift cut of z$ \geqslant 0.4$ complements previous WL work with galaxy clusters (which were mostly at redshifts $<0.3$, e.g \citealt{Okabe15, CCCP}), and the richness cut of $\lambda_{\rm{SDSS}} \geqslant 60$ has additional benefits: the SDSS photometric depth precludes a robust detection of redshift 0.6 clusters with lower richness, but a systematic spectroscopic follow-up using SPIDERS program only confirms $70\%$ of clusters at z$> 0.5$ with richness above 10, while all CODEX $\lambda_{\rm{SDSS}} \geqslant 60$ clusters are spectroscopically confirmed \citep{Clerc16}. It also precludes X-ray AGNs in low richness galaxy groups masquerading as high luminosity clusters.

\begin{figure}
\centering
 \includegraphics[width=9.5cm]{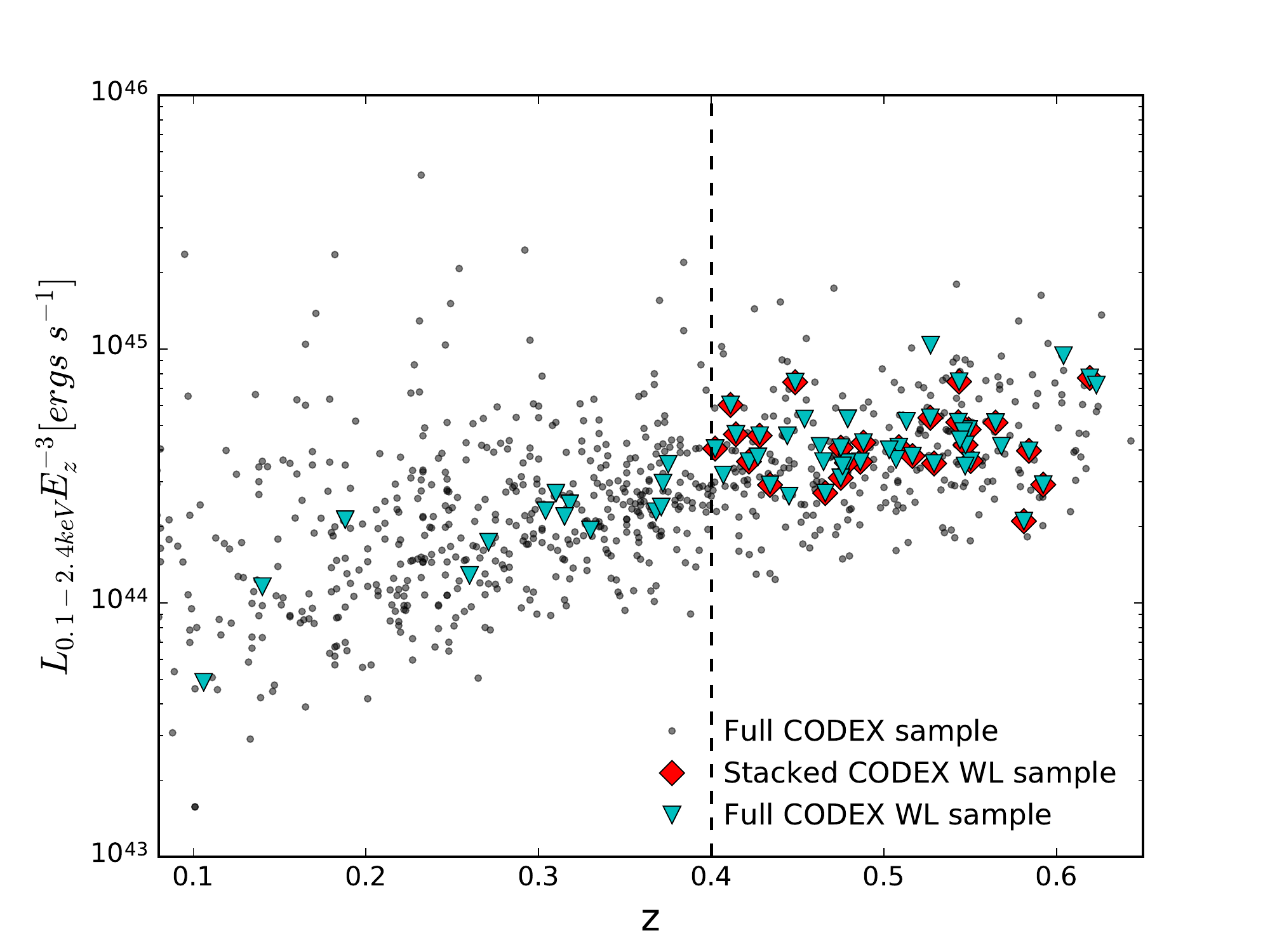}
 \caption{Distribution of corrected X-ray luminosity as a function of redshift for the full CODEX sample with $\lambda_{\rm{SDSS}} \geqslant 60$ clusters (black dots), for the full CODEX CFHT WL sample (cyan triangles) and for the 27 CFHT WL systems analysed in this work (red diamonds).}
\label{lx_full}
\end{figure}

\begin{figure}
\centering
 \includegraphics[width=9.5cm]{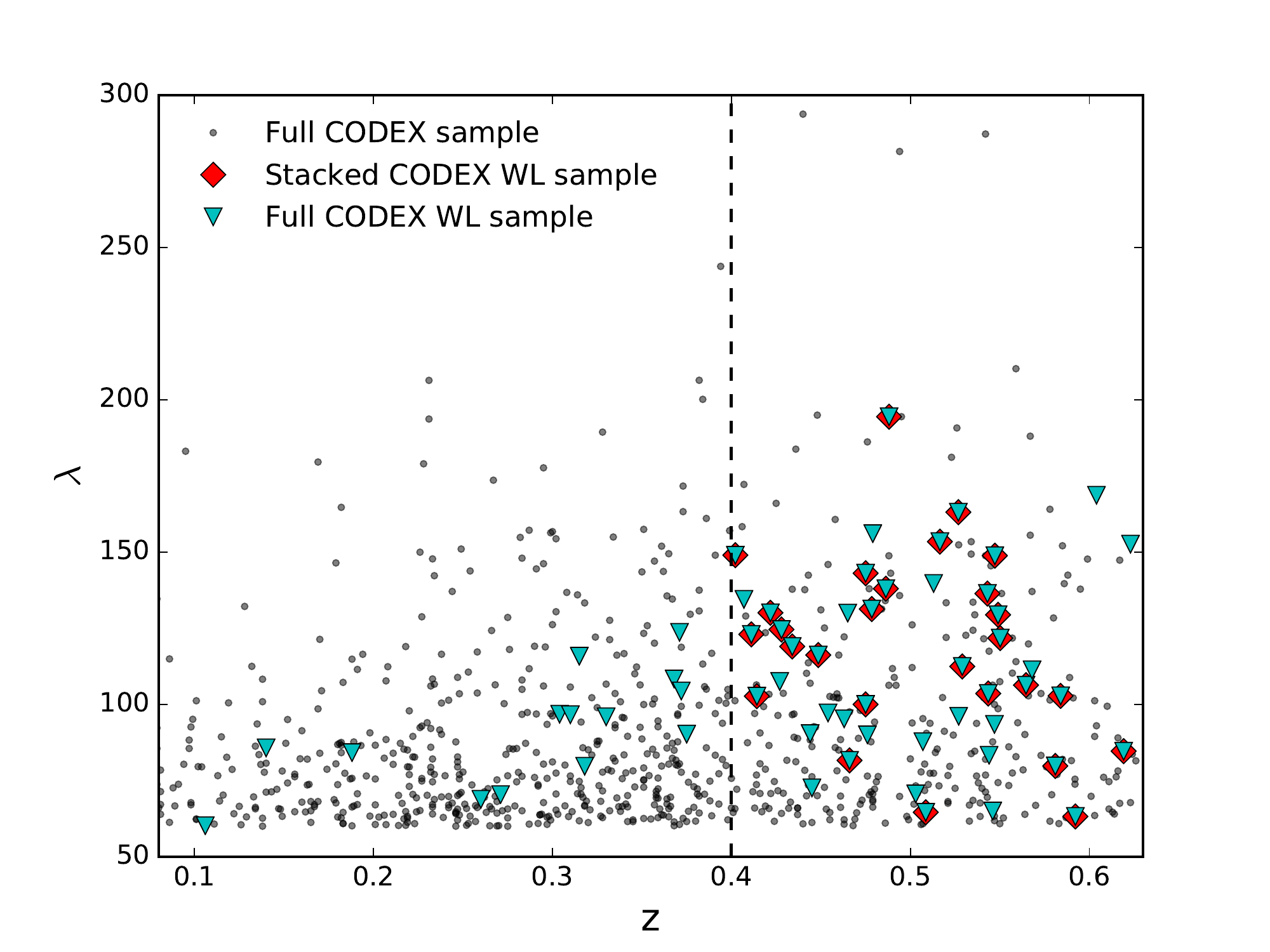}
 \caption{Distribution of richness  as a function of redshift for the full CODEX sample with $\lambda_{\rm{SDSS}} \geqslant 60$ clusters (black dots), for the full CODEX CFHT WL sample (cyan triangles) and for the 27 CFHT WL systems analysed in this work (red diamonds).}
\label{lambda_full}
\end{figure}

Figure \ref{lx_full} shows the X-ray luminosity distribution and figure \ref{lambda_full} the richness distribution, both for clusters with $\lambda_{\rm{SDSS}} \geqslant 60$, for the different samples as a function of redshift (spectroscopic when available). Grey dots represent the full high richness CODEX sample, cyan triangles the full CFHT WL sample (primary, secondary and tertiary with $\lambda_{\rm{SDSS}} \geqslant 60$) and red diamonds the 27 CFHT WL systems studied in this work. 

We quote the mean X-ray luminosity and richness with the corresponding uncertainties for the different samples in table \ref{tab_props}, computed for systems with $z \geqslant 0.4$, $\lambda_{\rm{SDSS}} \geqslant 60$. Based on this comparison there is no clear evidence that we are probing higher masses than the average high richness, high redshift CODEX sample, as there is no bias in L$_X$ but a small positive bias in $\lambda_{\rm{SDSS}}$. We note, however, that we did not apply any further richness cut to select our sample besides the initial $\lambda_{\rm{SDSS}} \geqslant 60$ one. 

\begin{table}
\centering
\caption{Samples properties for clusters with $z \geqslant 0.4$, $\lambda_{\rm{SDSS}} \geqslant 60$. }
\label{tab_props}
\begin{tabular}{|l | c c c|}
\hline
Sample  & $ N_{\rm{clust}}$ &  $\lambda_{\rm{SDSS}}$ & $L_{0.1-2.4keV} E_z^{-1}$ \\
& & & [10$^{44}$ ergs s$^{-1}$]  \\
\hline
All CODEX     & 298 & $ 97 \pm 2 $ &  $4.5 \pm 0.2$\\
All CFHT      & 46 & $113 \pm 4 $ &  $4.6 \pm 0.2$ \\
CFHT Stacked  & 27 & $119 \pm 6 $ &  $4.4 \pm 0.3$ \\
\hline
\end{tabular}
\end{table}

From the distribution of the X-ray luminosity in figure \ref{lx_full} we observe a decrease in the scatter with redshift: for $z < 0.4$ we find a scatter of $3.4 \times 10^{44}\ [ergs\ s^{-1}]$ and for $z \geqslant 0.4$ we find $2.6 \times 10^{44}\ [ergs\ s^{-1}]$. Based on this evolution of the spread in X-ray luminosity for a fixed richness cut we conclude that the likely explanation for the high mass we observe is the reduction of the scatter in the L$_X$-M$_{tot}$ relation with increasing redshift, which has also been suggested in other studies \citep{Mantz16a}. The physical interpretation of the decrease in the scatter requires a reduced contribution of cool cores to the total L$_X$.


\section{Data}
\label{data}

\ \ The present study is based on 25 fields with a total of 27 individual clusters from the CODEX CFHT WL sample. These are the fields with observations completed in the 'ugriz' bands, thus having a photometric redshift estimation for all the sources in the field. The systems have spectroscopic redshifts obtained from BOSS \citep{Clerc16} or NOT programs (section \ref{spec}), with the exception of one cluster that only has photometric redshift. Tables \ref{tab_cl} and \ref{tab_clx} detail these clusters.

\begin{table*}
\centering
\caption{CODEX weak lensing clusters sample.}
\label{tab_cl}
\begin{tabular}{l l l c c c c c c p{4.5cm}}
\hline
ID & RA$_{opt}$ & Dec$_{opt}$ & spec-z & photo-z$^a$ & $\delta$photo-z$^a$ & $\lambda_{\rm{SDSS}}^a$ & $\delta \lambda^a$ & $P^a_{CEN}$ & Previous name$^b$ \\

\hline
27974 & 00:08:51 & +32:12:25 & 0.474 & 0.491 & 0.014 & 100 & 25 & 0.91 & - \\
27940 & 00:20:08 & +34:51:18 & 0.454 & 0.460 & 0.011 & 116 & 24 & 1.00 & -  \\
55181 & 00:45:12 & -01:52:30 & 0.546 & 0.542 & 0.020 & 149 & 43 & 0.94 & ACT-CL J0045.2-0152 \\
64232 & 00:42:33 & -11:01:59 & - & 0.529 & 0.021 & 112 & 37 & 0.57 & -  \\
46649 & 01:35:17 & +08:47:49 & 0.619 & 0.536 & 0.024 & 85 & 31 & 0.99 & -  \\
59915 & 01:25:05 & -05:31:05 & 0.475 & 0.489 & 0.012 & 143 & 25 & 0.97 & -  \\
12451 & 08:04:39 & +53:25:44 & 0.577 & 0.566 & 0.026 & 103 & 35 & 0.95 & - \\
24865 & 08:22:42 & +41:27:29 & 0.483 & 0.477 & 0.012 & 138 & 23 & 0.79 & GMBCG J125.67646+41.45847 \\
24872 & 08:26:06 & +40:17:31 & 0.405 & 0.391 & 0.013 & 149 & 10 & 0.82 & WHL J082605.8+401731 \\
24877 & 08:24:27 & +40:06:18 & 0.588 & 0.539 & 0.021 & 63 & 59 & 0.49 & -  \\
24981 & 08:56:13 & +37:56:17 & 0.412 & 0.411 & 0.012 & 123 & 12 & 1.0 & WHL J085612.7+375615  \\
29283 & 08:04:35 & +33:05:10 & 0.554 & 0.536 & 0.019 & 129 & 30 & 0.90 & WHL J080433.5+330511  \\
29284 & 08:03:30 & +33:01:48 & 0.550 & 0.557 & 0.023 & 122 & 33 & 0.70 & WHL J080329.8+330146  \\
43403 & 08:10:18 & +18:15:18 & 0.422 & 0.418 & 0.012 & 130 & 10 & 0.96 & GMBCG J122.57586+18.25506 \\
47981 & 08:40:03 & +08:37:55 & 0.543 & 0.551 & 0.023 & 136 & 33 & 0.98 & -  \\
52480 & 09:34:39 & +05:41:46 & 0.561 & 0.546 & 0.022 & 106 & 54 & 0.70 & WHL J093439.0+054144 \\
25424 & 11:30:56 & +38:25:08 & 0.503 & 0.513 & 0.017 & 65 & 17 & 0.73 & WHL J113100.4+382437  \\
29811 & 11:06:08 & +33:33:40 & 0.487 & 0.495 & 0.013 & 194 & 31 & 1.0 & WHL J110608.5+333339  \\
25953 & 14:03:44 & +38:27:04 & 0.477 & 0.484 & 0.012 & 131 & 19 & 0.78 & WHL J140344.1+382703  \\
35361 & 14:56:11 & +30:21:04 & 0.415 & 0.411 & 0.012 & 103 & 9 & 0.66 & WHL J145611.1+302104 \\
35399 & 15:03:03 & +27:54:58 & 0.519 & 0.534 & 0.018 & 153 & 31 & 0.56 & WHL J150303.7+275519  \\
36818 & 22:20:16 & +27:20:02 & 0.582 & 0.578 & 0.030 & 80 & 64 & 0.62 & -  \\
37098 & 23:19:17 & +28:12:00 & 0.544 & 0.573 & 0.028 & 104 & 35 & 0.97 & -  \\
41843 & 23:40:45 & +20:52:05 & 0.432 & 0.435 & 0.011 & 119 & 13 & 0.59 & -  \\
50492 & 23:32:14 & +10:36:36 & 0.524 & 0.524 & 0.017 & 163 & 30 & 0.92 & -  \\
50514 & 23:16:43 & +12:46:55 & 0.468 & 0.463 & 0.011 & 82 & 13 & 0.89 & -  \\
54795 & 23:02:16 & +08:00:29 & 0.431 & 0.429 & 0.011 & 125 & 35 & 0.76 & WHL J230208.1+080118 \\
\hline
\end{tabular}
\begin{tabular}{p{20cm}}
$^a$ Parameters assigned by redMaPPer\\
$^b$ In cases where there is more than one possible match we listed here only the nearest object. 
\end{tabular}
\end{table*}

\begin{table}
\caption{X-ray properties of the CODEX weak lensing clusters sample.}
\label{tab_clx}
\begin{tabular}{p{1.3cm} p{1.7cm} p{1.7cm} p{2.1cm}}
\hline
ID &  RA$_{X}$ & Dec$_{X}$ & L$_X^a$\ \ \ \ \ \ \ \ \ \ [$10^{44}$erg s$^{-1}$] \\

\hline
27974 & 00:08:58 & +32:12:04  & 8.1 $\pm$ 3.6 \\
27940 & 00:20:10 & +34:53:36 & 9.2 $\pm$ 2.1 \\
55181 & 00:45:10 & -01:51:49 & 5.9 $\pm$ 2.3 \\
64232 & 00:42:32 & -11:04:07 & 4.8 $\pm$ 1.8 \\
46649 & 01:35:17 & +08:48:14 & 14.2 $\pm$ 4.6 \\
59915 & 01:25:01 & -05:31:53 & 3.9 $\pm$ 1.4 \\
12451 & 08:04:38 & +53:25:24 & 5.9 $\pm$ 2.0 \\
24865 & 08:22:45 & +41:28:09 & 4.9 $\pm$ 1.7 \\
24872 & 08:25:59 & +40:15:19 & 5.4 $\pm$ 1.3 \\
24877 & 08:24:40 & +40:06:53 & 4.9 $\pm$ 2.0 \\
24981 & 08:56:14 & +37:55:52 & 7.6 $\pm$ 1.9 \\
29283 & 08:04:36 & +33:05:27 & 7.0 $\pm$ 2.3 \\
29284 & 08:03:30 & +33:02:06 & 4.9 $\pm$ 1.9 \\
43403 & 08:10:20 & +18:15:13 & 4.7 $\pm$ 1.7 \\
47981 & 08:40:02 & +08:38:04 & 6.9 $\pm$ 2.6 \\
52480 & 09:34:37 & +05:40:53 & 7.6 $\pm$ 2.4 \\
25424 & 11:31:01 & +38:24:42 & 5.5 $\pm$ 2.1 \\
29811 & 11:06:07 & +33:33:36 & 5.6 $\pm$ 1.8 \\
25953 & 14:03:42 & +38:27:38 & 4.6 $\pm$ 1.3 \\
35361 & 14:56:13 & +30:21:12 & 6.0 $\pm$ 1.3 \\
35399 & 15:03:10 & +27:55:00 & 4.8 $\pm$ 1.8 \\
36818 & 22:20:21 & +27:21:00 & 3.0 $\pm$ 1.5 \\
37098 & 23:19:16 & +28:12:01 & 10.0 $\pm$ 4.4 \\
41843 & 23:40:45 & +20:53:02 & 3.7 $\pm$ 1.4 \\
50492 & 23:32:14 & +10:35:32 & 7.4 $\pm$ 2.2 \\
50514 & 23:16:46 & +12:47:12 & 3.6 $\pm$ 1.3 \\
54795 & 23:02:17 & +08:02:14 & 5.9 $\pm$ 1.6 \\
\hline
\end{tabular}
$^a$X-ray luminosity is given for the energy range 0.1 - 2.4 keV, computed within R$_{500}$.
\end{table}

\subsection{Data Reduction}

\ \ To process the CODEX data, we use the algorithms and processing
pipelines (\textit{theli}) developed within the CFHTLS-Archive
Research Survey \citep[CARS; see][]{Erben09, Erben05, Schirmer13} and
the CFHT Lensing Survey (CFHTLenS; see \citealt{Heymans12, Hildebrandt12, Erben13} and http://cfhtlens.org).
CFHTLenS and CODEX have similar observing strategies (five-band
multi-colour data for photometric redshifts) and science goals (weak
lensing studies). This allowed a direct transfer of our CFHTLS
expertise to CODEX. In the following we only give a very short
description of our procedures to arrive at the final co-added images
for CODEX. All algorithms and prescriptions are described in detail in \citet{Erben13}, which an interested reader should consult as well as the references therein. Below, we also give a more detailed analysis of the quality from our photometric calibration which is crucial for the quality of photometric redshift estimates.

\begin{figure}
\centering
  \includegraphics[width=8.6cm]{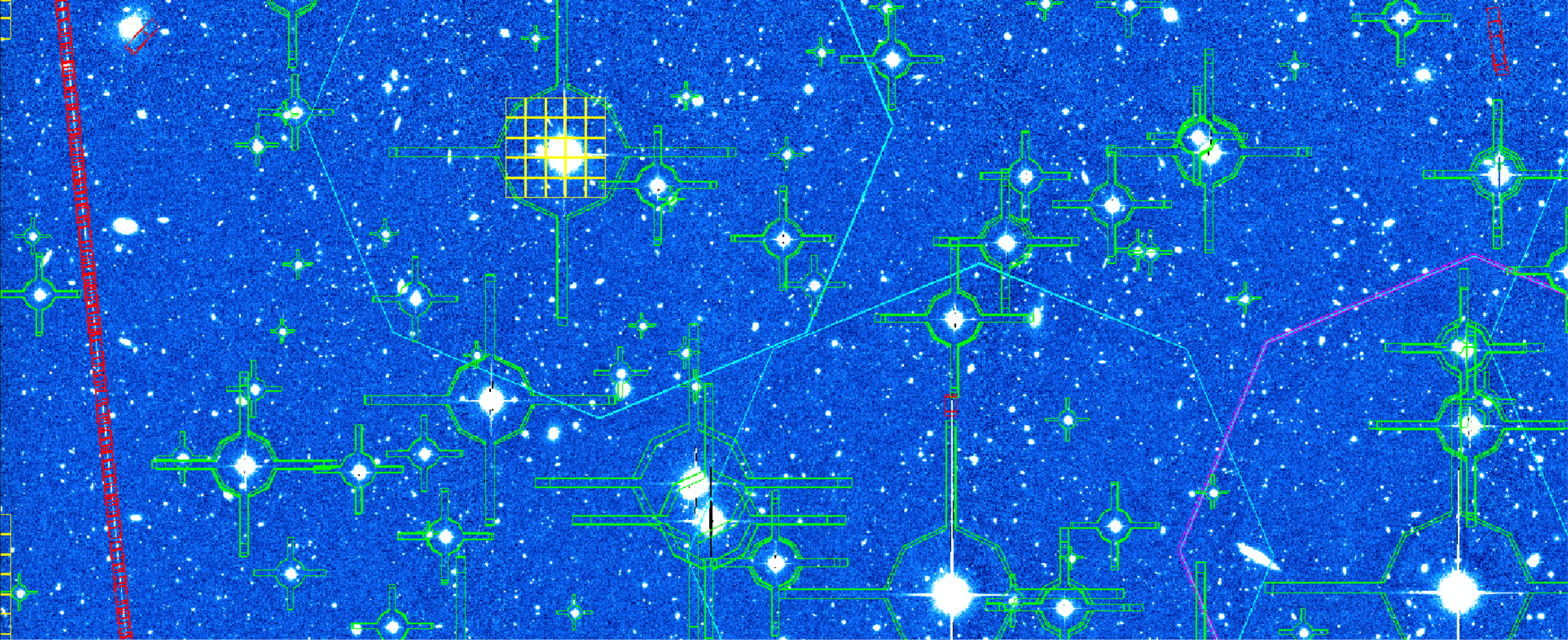}
  \caption{Detail of image masks for the reduced science frame of CODEX 12451 (\textit{i}' band). Bad area masks are shown in green, yellow, magenta, cyan and red.}
\label{mask}
\end{figure}

Our CODEX data processing consisted of the following steps:
\begin{enumerate}
\item Data sample:\\
  We start our data analysis with the
  \textit{Elixir}\footnote{\url{http://www.cfht.hawaii.edu/Instruments/Elixir/}}
  preprocessed CODEX data available at the Canadian Astronomical Data
  Centre
  (CADC)\footnote{\url{http://www4.cadc-ccda.hia-iha.nrc-cnrc.gc.ca/cadc/}}.

  For the current study we used CODEX observations obtained from 2012 until 2015. The data were obtained under several CFHT programs (PI J.P. Henry, 12AH24, 12AH99, 12BH07, 13AH17, 13AH99, 15AH81; PI L. van Waerbeke, 12AC26,  12BC19, 13AC10; PI R. Dupke, 12AB16, 12AB99, 12BB06, 13AB01, 13BB03, 13BB99; PI J.P. Kneib, 12BF11, 13AF12, 14BF11;  PI A. Finoguenov, 14BC19).   

\item Single exposure processing:\\
  The \textit{Elixir} preprocessing includes a complete removal of the
  instrumental signature from raw data \citep[see also][]{Magnier04}.
  In addition, each exposure comes with all necessary photometric
  calibration information. Therefore, we only need to perform the
  following processing steps on single exposures: (1) we identify and
  mark individual exposure chips that should not be considered any
  further. This mainly concerns chips which are completely dominated
  by saturated pixels from a bright star. (2) We create sky-subtracted
  versions of the images. In addition, we create a weight image for each
  science chip. It gives information on the relative noise properties
  of individual pixels and assigns a weight of zero to defective
  pixels (such as cosmic rays, hot and cold pixels, areas of satellite
  tracks). (3) We use \textsl{SExtractor} to extract high $S/N$
  sources\footnote{We consider all sources having at least 5 pixels
    with at least $5\sigma$ above the sky-background variation.} from
  the science image and weight information. These source catalogues are
  used to astrometrically and photometrically calibrate the data in
  the next processing step. In addition we perform an analysis of the
  PSF anisotropy and use this information to reject images showing
  high stellar ellipticities. Highly elongated point sources are a
  good indication of tracking issues or other severe problems during
  an exposure.
\item Astrometric and Photometric calibration: \\
  We use the \textsl{scamp}
  software\footnote{\url{http://www.astromatic.net/software/scamp}}
  \citep[see][]{Bertin06} to astrometrically calibrate the CODEX
  survey. We use the Two Micron All Sky Survey
  \citep[2MASS;][]{Skrutskie06} as astrometric reference and calibrate
  simultaneously all filters from a specific CODEX-pointing with
  \textsl{scamp}. Once an astrometric solution is established we use
  overlap sources from individual exposures to establish an internal,
  relative photometric solution for all exposures. We reject all
  exposures with an absorption of more than 0.2
  magnitudes
  \footnote{Our rejection limit of 0.2
    magnitudes turned out to be a very good conservative limit to
    reject the small fraction of images observed under poor
    photometric conditions (see Sec. 3.3 (iii) of \citealt{Erben13}).} and rerun \textsl{scamp} on the remaining
  images. With the relative photometric solution and the
  \textit{Elixir} zeropoint information we estimate a
  photometric zeropoint for each filter.
\item Image co-addition and mask creation: \\
  The next step of our image processing co-adds the sky-subtracted
  exposures belonging to a pointing/filter combination with the
  \textsl{SWarp}
  program\footnote{\url{http://www.astromatic.net/software/swarp}}
  \citep[see][]{Bertin02}. The stacking is performed with a
  statistically optimally weighted mean which takes into account
  sky-background noise, weight maps and the \textsl{scamp} relative
  photometric zeropoint information. As a final step we use the
  \textit{automask} tool \citep[see][]{Dietrich07} to create image
  masks for all pointings. These masks flag bright, saturated stars
  and areas which would influence the analysis of faint background
  sources. Figure \ref{mask} shows a typical reduced science frame including bad area masks. On average the masks represent $25\%$ of the total area and for the worst case, the field CODEX 36818, it corresponds to $43\%$ of the area.
\end{enumerate}
    
Table \ref{tab:obs} in the appendix provides observational information about the fields used in this work. The limiting magnitude $m_{lim}$ is the $5\sigma$ limit in a 2" radius aperture (as defined in Sec. 2 of \citealt{Erben13}).

\subsection{Photometric Catalogues}
\label{photoz}

\ \ The photometric multi-band catalogue creation and photometric redshift (photo-z) estimation are performed as described in \citet{Brimioulle13}. We only give a brief overview here. 

In the first step we match the point spread function (PSF) of the different filter observations to allow reasonable aperture colour estimates for the catalogue objects. For that we degrade all filter images to the seeing of the worst band (in general the u$^{*}$-band), convolving them with a global Gaussian kernel. We run the \textsl{SExtractor} software in dual-image mode, using the unconvolved  i-band image as detection band and the convolved images for flux and magnitude extractions, including weight images and masks of bad area. We extract all objects which are at least 2$\sigma$ above the background on four contiguous pixels.

Zeropoint and extinction corrections are performed by stellar locus regression using the stellar library of \citet{Pickles98} as a reference. We then estimate photo-z's using the template-fitting code of \citet{Bender01}. The used template set includes templates developed in \citet{Bender01}, 13 SEDs from \citet{Ilbert13} and linear combinations.

\subsection{Shear measurement}
\label{lensfit}

\ \ We measure galaxy shapes with the \textsl{lensfit} algorithm \citep{Miller13} using the i-band images. We choose the i-band because it detects a  galaxy population with the highest mean redshift, and thus lensing signal, for a fixed exposure time among all filters and produces some of the best quality images from our data-set.

The quantities given by \emph{lensfit} and used in the analysis are the measured ellipticity components $e_1$ and $e_2$ and the weight \emph{w}. The lensing weight \emph{w} accounts for both the ellipticity measurement error and for the intrinsic shape noise and is defined in \citet{Miller13}.

We filter for sources with reliable shape measurements by applying a cut for objects with \emph{lensfit weight}  $w > 0$ and \emph{lensfit fitclass} $= 0$.
 
The version of \emph{lensfit} used is this work is the new ‘self-calibrating’ version presented in \citet{Fenech16}. We refer the reader to \citet{Fenech16} and to its first use in \citet{Hildebrandt16} for the details, but it is worth highlighting a few important facts about the self-calibration. Its main purpose is to resolve the noise bias problem plaguing shape measurements techniques (e.g. \citealt{Melchior12, Refregier12, Miller13}). The correction however is not perfect, it was shown to contain residual calibration bias of the order of $2\%$. In \citet{Fenech16}, it was discussed how this could be reduced further, with the aid of image simulations, in order to reach the sub-percent goal for the cosmic shear analysis presented in \citet{Hildebrandt16}. For our purpose here, that is the lensing analysis of the CODEX clusters, the residual statistical error is of the order of 10-15 percents, which significantly exceeds the self-calibrating \emph{lensfit} accuracy. We therefore decided to ignore the calibration refinement discussed in \citet{Fenech16}, and work with the self-calibrated \emph{lensfit} shapes only. It should be noted, however, that the shear calibration error could still substantially depend on the source galaxy size and magnitude. Such dependence could be hidden in the source selection criteria discussed in Section \ref{back_sel} and Appendix \ref{ap:photoz}. How the results depend on the selection criteria is discussed in \ref{conclusion} and Appendix \ref{ap:photoz}.

\subsection{Spectroscopic follow-up}
\label{spec}

\ \ A weak lensing analysis requires precise knowledge of the lens redshift. Therefore additional CODEX spectroscopy has been carried out using the Nordic Optical Telescope (NOT). These observations were obtained from 2014 to 2016 under several NOT programs (PI A. Finoguenov, 48-025, 51-035, 52-026, 53-020). 

Each cluster was observed in multi-object spectroscopy mode, targeting $\sim 20$ member galaxies including BCGs and having spectral resolving power of $\sim 500$. The BCG for each cluster was observed through a 1.5" slit for 2700 sec with a grism that provides wavelength coverage between approximately 400-700 nm. The average seeing over the four programs is near 1''. 

Because we are solely interested in the redshift of the Ca H+K lines, only wavelength calibration frames are additionally obtained. Standard \textsl{IRAF}\footnote{\url{http:iraf.noao.edu}} packages are used in the data reduction, spectra extraction, and wavelength calibration process. The redshifts are determined finally using {\it rvidlines} to measure the positions of the two calcium lines for a weighted average fit.

\subsection{Background selection}
\label{back_sel}

\ \ The weak lensing signal from a system composed by a lens and a source is proportional to the surface mass density of the lens and to the angular diameter distances between observer and lens ($D_{\rm{d}}$), and the ratio $\beta$ of the distance between lens and source ($D_{\rm{ds}}$) and observer and source ($D_{\rm{s}}$):

\begin{equation}
\beta = \frac{D_{\rm{ds}}}{D_{\rm{s}}}
\end{equation}

As a consequence, in order to constrain the lens properties it is necessary to know the redshift distribution of the sources and to exclude the contamination of the background sample with cluster members.


We select the background population and estimate its redshift distribution based on the multi-band information available for the fields. Our primary method is a colour-magnitude decision tree, described in the following. We also applied a photometric redshift point estimator cut described in appendix \ref{ap:photoz}, where we compare the results from the two different methods.


The colour-magnitude decision tree estimates the redshift distribution of each galaxy in the cluster fields by comparison of our multi-band photometric data to a reference catalogue of high-quality photometric redshifts from CFHTLS Deep\footnote{http://www.cfht.hawaii.edu/Science/CFHLS/ \\ cfhtlsdeepwidefields.html}. The method is described in detail in \citet{Gruen16}. We only give a brief summary here.

To assign a $p(z)$ to each source galaxy, we split the $u^{\star}, g', r',i', z'$ colour-magnitude space into boxes (hyperrectangles), respecting the strict magnitude limit of $i'<24.7$ of the shape catalogue. The split conditions are chosen to maximise the lensing signal-to-noise ratio with a minimum of 1000 galaxies in the CFHTLS Deep fields in each box (cf. \citealt{Gruen16}, their Section 2.1 and Appendix A). For each source in our cluster fields, we take the observed distribution of redshifts of reference galaxies in the box it falls into as its $p(z)$.

Consequently, for a galaxy $j$ in the cluster field falling into a colour-magnitude box $B_i$, we estimate its $\beta_j$ as the mean of all reference galaxies in $B_i$, i.e.

\begin{equation}
\beta_j=\langle\beta\rangle^{corr}_{B_i} \; .
\label{betacm}
\end{equation}

\noindent This includes a noise bias correction as follows. Galaxies are weighted in the lensing analysis by their estimated $\beta$ (eq. \ref{delsig} and \ref{scrit}); because galaxies for which we overestimated $\beta$ are weighted up relative to galaxies for which we underestimated $\beta$, there is a small noise bias when $\sigma_{\beta,B_i}>0$, which we correct by

\begin{equation}
\langle\beta\rangle^{corr}_{B_i} = \langle\beta\rangle_{B_i}/(1+\sigma^2_{\beta,B_i}/\langle\beta\rangle_{B_i}^2) \; .
\end{equation}

\noindent where we estimate the cosmic variance based uncertainty of the mean $\beta$ in each colour-magnitude box by a jackknife over the four pointing of CFHTLS Deep, i.e. as

\begin{equation}
\sigma_{\beta,B_i}= \sqrt{\frac{3}{4}\sum_{j=1}^4 [ (\langle\beta\rangle_{B_i,\neg j}-\langle\beta\rangle_{B_i})^2 ] } \; ,
\end{equation}

with $\langle\beta\rangle_{B_i,\neg j}$ being the estimate of the lensing-weighted mean $\beta$ of the reference sample in $B_i$ when excluding CFHTLS Deep pointing $j$, and $\langle\beta\rangle_{B_i}$ is the average over all $\langle\beta\rangle_{B_i,\neg j}$.

This correction is accurate for Gaussian statistical uncertainty on $\langle\beta\rangle_{B_i}$, a justified assumption due to the large number of reference galaxies in each box.

\begin{figure}
\centering
 \includegraphics[width=8cm]{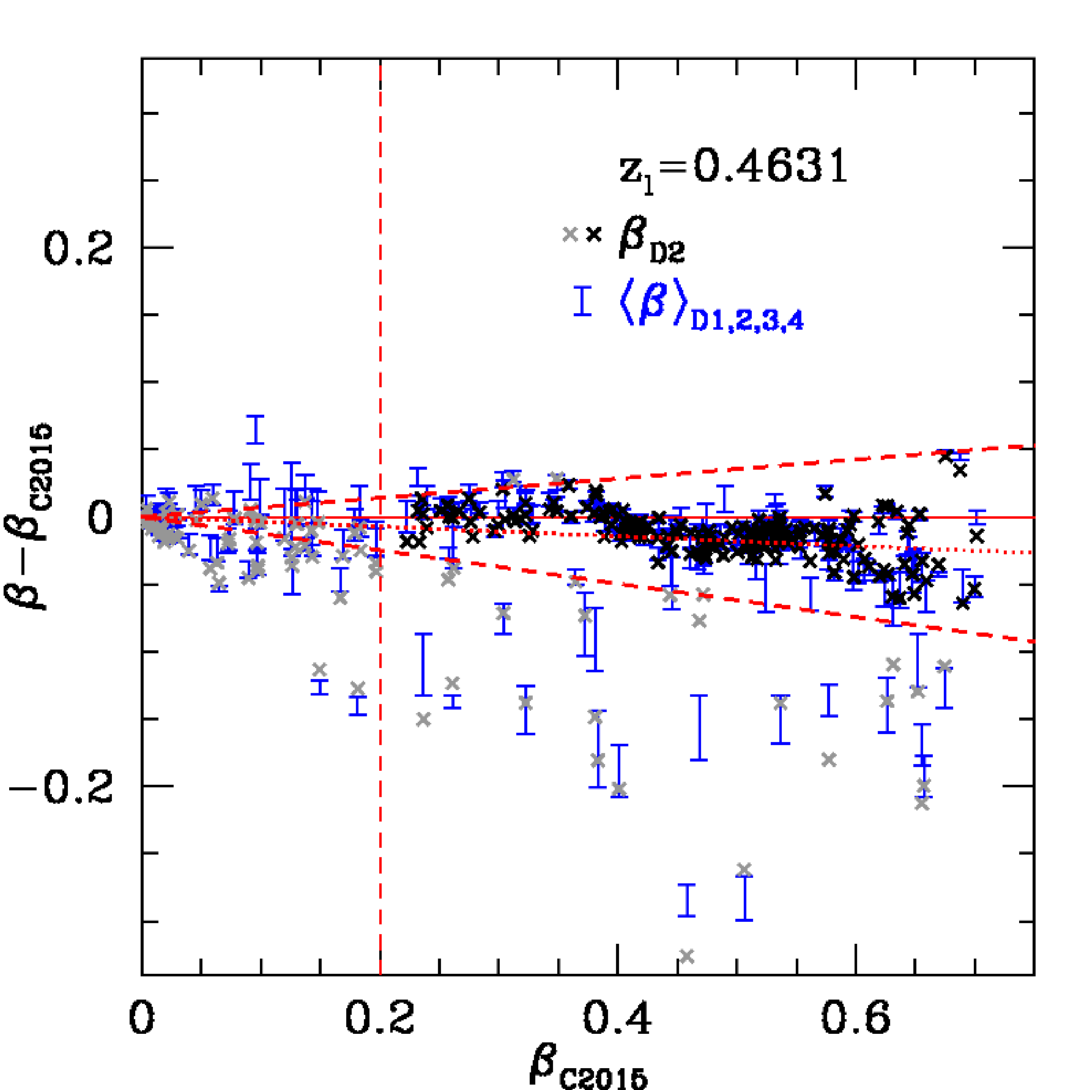}
 \caption{Mean value of $\beta$ in boxes in 5-band colour-magnitude spaces estimated from COSMOS2015 photo-z ($\beta_{C2015}$), matched galaxies with CFHT Deep + WIRDS photometry ($\beta_{D2}$, black / grey points), and CFHT Deep + WIRDS galaxies in all four Deep pointings (blue error bars indicating jackknife uncertainty). Dotted red line indicates the median ratio of $\beta_{D2}/\beta_{C2015}$. Dashed red lines marks exclusion criteria of $\beta_{C2015}<0.2$ or 10$\%$ deviation from this median ratio, outside of which leaves are excluded from the source sample (grey points). Results are shown for a lens redshift of $z_l=0.4631$ (CODEX50514).}
\label{deadleaf}
\end{figure}

As an additional reference catalogue, we use the COSMOS2015 sample by \citet{Laigle16}, who find excellent metrics ($\sigma_{\Delta z/(1+z)} = 0.007$ at $I < 22.5$ and $0.021$ for high-$z$, blue galaxies, yet with a substantial outlier fraction). We cross-match this catalogue to CFHTLS D2 objects for direct validation of our $\langle\beta\rangle$ estimates, applying the following cuts to select the sample of sources:

\begin{enumerate}
\item We remove all galaxies from the cluster and reference fields whose colour is in the range spanned by galaxies in the reference catalogue best fitted by a red galaxy template in the redshift interval $z_d\pm0.04$. This filter is applied \emph{before} constructing the decision tree.
\item We remove galaxies in colour-magnitude boxes for which the estimate of $\langle\beta\rangle$ from COSMOS2015 redshifts is below 0.2 (fig. \ref{deadleaf}).
\item We remove galaxies in colour-magnitude boxes populated with any galaxies in the reference catalogue for which the redshift estimate is within $z_d\pm0.06$ to prevent contamination of the source sample with cluster members.
\item We remove galaxies in colour-magnitude boxes for which the estimates of $\langle\beta\rangle$ in a matched catalogue of CFHTLS D2 and COSMOS2015 redshift deviate by more than 10\% from the median ratio over all boxes (fig. \ref{deadleaf}).
\end{enumerate}

\noindent Because we require colour-magnitude boxes to contain at least 1000 reference galaxies, the latter two cuts cause no significant bias against densely populated parts of colour-magnitude space.  

We expect this method to deliver a more accurate characterisation of the background population than individual point estimates as: 
\begin{itemize}
\item it is based on more accurate photometric redshifts from deeper reference catalogues with more bands (NIR)
\item we have excluded spurious regions in the colour-magnitude space by comparing two different reference photo-z estimates;
\item the full PDF information within a colour-magnitude region is used, with the empirical distribution of galaxies in CFHTLS Deep at the same color-magnitude as our source galaxies as a prior;
\item we have consecutively excluded regions of colour-magnitude space populated by galaxies at the cluster redshift.
\end{itemize}

Before applying the background selection based on the colour-magnitude method we had a total number of 1,242,369 galaxies with a shear weight \emph{w} $> 0$. Applying criteria \textit{ii} and \textit{iv} leads to a background population corresponding to $60\%$ of the initial sample. A background selection based on criterion \textit{iii} gives a subsample with $42\%$ the size of the initial one. From these numbers the criterion \textit{iii} has shown to be the most restrictive one. We adopt all criteria (\textit{ii}, \textit{iii} and \textit{iv}) to select the sources, leading to a final subsample with $38\%$ the size of the initial sample. We note that a large fraction of the initial sample of galaxies have a small expectation value for $\beta$ and would have a small contribution to the final shear signal.


We decided to keep the most conservative cut to lower the systematic uncertainties arising from cluster member contamination at the price of higher statistical uncertainty. This conservative selection also results in a lower density of sources in comparison to the photometric redshift point estimator cut ($\sim 40\%$ lower), reported in apendix \ref{ap:photoz}. 

The total weighted number density of background sources for the colour-magnitude selection method, after stacking the 27 individual clusters, is 120 gal/arcmin$^2$, given by:

\begin{equation}
    N = \frac{1}{\rm{A}} \frac{(\sum w_i)^2}{\sum w_i^2}
\label{total_num}
\end{equation}

\noindent where \emph{w$_i$} is the individual shear weight of each background galaxy and \emph{A} is the area of a circle of radius $15$ arcmin$^2$ within which we compute the sources. 

We also determine the effective weighted number density of galaxies following \citealt{Heymans12}:

\begin{equation}
    N = \frac{1}{\Omega} \frac{(\sum w_i)^2}{\sum w_i^2}
\label{effec_num}
\end{equation}

\noindent considering the total area of the fields excluding masked areas, $\Omega$. The effective number density of galaxies selected as background by the colour-magnitude method is 6 gal/arcmin$^2$.

\subsection{Errors}
\label{errors}

In this section we describe the systematic uncertainties arising from shear measurements and redshift estimates, as the colour-magnitude decision tree method allows to quantify the systematics. We assimilate both shear and redshift errors in the Bayesian analysis (section \ref{fit}) when fitting the model to data.     

\subsubsection{Shear systematics}

\ \ As mentioned in Section \ref{lensfit}, we expect the residual uncertainty on the \emph{lensfit} shapes to be of the order of $~2$ percent according to previous results \citep{Fenech16}. This systematic uncertainty is taken into account in our modelling by introducing a shear calibration factor with mean $\delta_{s}=0$ and Gaussian width $\sigma_{s}=0.02$.

\subsubsection{Colour-magnitude decision tree systematics}

\ \ Uncertainties on the colour-magnitude decision tree method arise from cosmic variance and errors in the reference catalogue photometric redshifts. The photo-z uncertainties are assessed by comparing the values of $\beta$ obtained from the CFHTLS D2 field ($\beta_{D2}$) and COSMOS2015 ($\beta_{DC2015}$) catalogues. The difference on $\beta$ from this matched catalogue is free of cosmic variance as we apply two different template fits over the same galaxies. We compute the mean shift for each individual cluster \emph{i} as 

\begin{equation}
\delta_{\rm{cm},i} = \frac{1}{2} \frac{\left \langle \beta_{C2015} \right \rangle - \left \langle \beta_{D2} \right \rangle}{\left \langle \beta_{D2} \right \rangle}
\end{equation}

\noindent and the variance, assuming a Gaussian of the same variance as a top-hat distribution between $\beta_{D2}$ and $\beta_{DC2015}$, as

\begin{equation}
\sigma_{\rm{cm},i} = \frac{1}{\sqrt{3}} \lvert \delta_{\rm{cm, i}} \rvert
\end{equation}

The cosmic variance contribution $\sigma_{\rm{cv}}$ of the mean $\beta$ of each cluster is calculated by a jackknife estimate over the four pointings of CFHTLS Deep, i.e. as

\begin{equation}
\sigma_{\rm{cv}, i} = \sqrt{\frac{3}{4}\sum_{j=1}^4 [ (\langle \beta_i \rangle_{\neg j}-{\langle \beta_i \rangle)^2 ] }/{\langle \beta_i \rangle}} \; ,
\end{equation}

where $\langle\beta_i\rangle_{\neg j}$ is the estimate of the lensing-weighted mean $\beta$ of the source sample around the cluster when excluding CFHTLS Deep pointing $j$, and $\langle\beta_i\rangle$ is the average over all pointings. 

To model the uncertainties on the stacked signal, we first assign a weight to each of our clusters:

\begin{equation}
\label{wi}
w^{\rm{cl}}_i = \frac{\sum_j (w_{j,i} \beta_{j,i})}{\sum_{k} \sum_{j} (w_{j,k} \beta_{j,k})}
\end{equation}

\noindent where \emph{j} runs over all galaxies in a cluster field, \emph{k} runs over all the clusters in our sample and \emph{w} is the lensing weight. 

We conservatively assume that the systematic uncertainties in redshift are fully correlated between the clusters and therefore take their weighted mean. For this we use the individual cluster weights (eq. \ref{wi}) to determine the mean shift $\delta_{\rm{cm}}$ and variance $\sigma_{\rm{cm}}$ over all the clusters:

\begin{equation}
\label{mumg}
\delta_{\rm{cm}} = \sum_i w^{\rm{cl}}_i \delta_{\rm{cm},i}
\end{equation}

\begin{equation}
\sigma_{\rm{cm}} = \sum_i w^{\rm{cl}}_i \sigma_i
\end{equation}

\noindent where the variance $\sigma_i$ is calculated taking into account both cosmic variance and photo-z uncertainties:

\begin{equation}
\label{sigsig}
\sigma_{i} = \sqrt{\sigma_{\rm{cv}, i}^2 + \sigma_{\rm{cm}, i}^2}
\end{equation}

We find for our sample $\delta_{\rm{cm}} = -0.016$ and $\sigma_{\rm{cm}} = 0.023$. 

The shear and redshift systematics enter the Bayesian analysis through a multiplicative factor $S_m$ which changes the amplitude of the theoretical density profile to assimilate these uncertainties. This factor is a free parameter with a prior defined as a normal distribution. If we define $\delta_{Sm} = 1 - \delta_{\rm{cm}} - \delta_{\rm{s}}$ and $\sigma_{Sm} = \sqrt{\sigma_{\rm{cm}}^2 + \sigma_{\rm{s}}^2}$, the parameter $S_m$ is modelled as:

\begin{equation}
\label{Sm}
ln (S_m) \propto \frac{(x - \delta_{Sm})^2}{2 \sigma_{Sm}^2}\ ,\ 1 - 5\sigma_{Sm} < x < 1 + 5\sigma_{Sm} 
\end{equation}

\section{Weak lensing Analysis}
\label{wl}

\ \ In this work we probe the gravitational potential of a sample of clusters by fitting their combined (stacked) radial shear profile. We adopt a model which describes the observed signal as a combination of the cluster signal (the main halo component), assumed to follow a NFW profile, the neighbouring halos (the two-halo term) and uncertainties related to the choice of the mass distribution centre (the off-centring formalism).

In the following sections we describe how we define the observed shear profile and the details of the adopted model. Section \ref{fit} describes the Bayesian analysis used to constrain our profile parameters and section \ref{results} present the results. In section \ref{compare} we compare our results with previous observational studies and with predictions from numerical simulations.

\subsection{Stacking}
\label{stack}

\ \ Triaxiality and substructure of dark matter halos plus the line-of-sight distribution of matter are systematic contributions that can affect weak lensing data. A common strategy to tackle these issues is combining several objects with similar properties, in an approach known as stacking \citep{Johnston07, Okabe10, Umetsu11a, Okabe13}, thus tending to average out those sources of bias. It also increases greatly the signal-to-noise ratio, as we multiply the density of background sources by the number of clusters on the stack. The downside of this technique is that it averages also the properties of the clusters. This is not a major concern in our study given that we seek to characterise global properties of the cluster population. 

Our cluster sample lies in the redshift range $0.40 < z < 0.62$ and has $\lambda_{\rm{SDSS}} \geqslant 60$. This richness cut excludes galaxy groups hosting X-ray AGN. We do not perform any additional selection on the X-ray luminosity but the sensitivity of RASS data, on which detection of clusters is based, leads to a selection close to $3 \times 10^{44}$ ergs s$^{-1}$ at the redshifts of interest.

We stack the shear signal in radial bins as a function of separation in physical units. The investigated radial range is $0.1$ Mpc h$^{-1}$ $<$ r $< 2.3$ Mpc h$^{-1}$ divided in 10 log-spaced bins. We define the centre of each cluster as the galaxy with the highest value of $P_{CEN}$ given by redMaPPer. This parameter is the probability that a given galaxy is the central galaxy (CG) of a cluster, taking into account the redshift and luminosity of the CG and of the cluster and the local red galaxies density. 

\begin{figure}
\centering
\includegraphics[width=8cm]{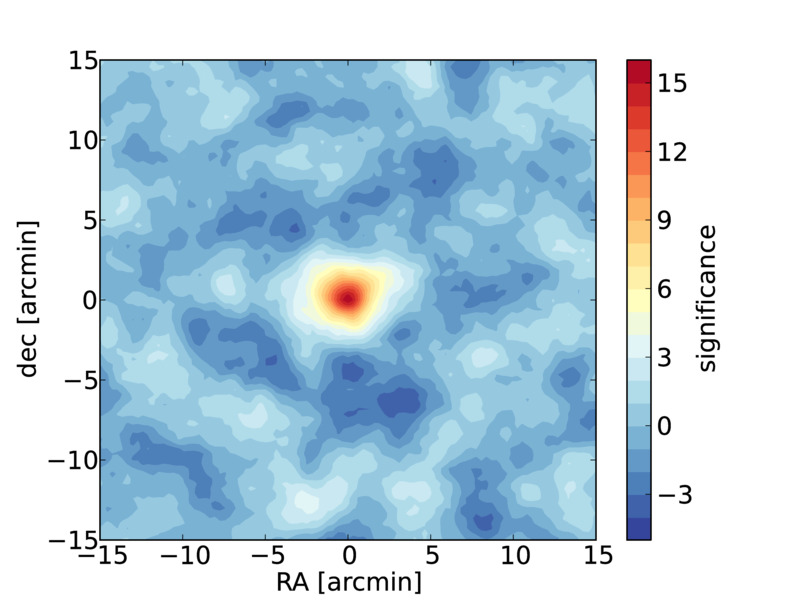}
 \caption{Aperture mass significance maps for the stacked cluster.}
\label{fig:massmap}
\end{figure}

In order to obtain a first two-dimensional view of the gravitational lensing signal, we calculate the so-called aperture mass map \citep{Schneider96} of our stacked galaxy cluster sample (see also \citealt{Gruen13}). To calculate the significance we use the Gaussian weight function

\begin{align}
  w_{\rm{g}}(\theta) \propto
  \begin{cases}
    \rm{exp}(-\vert \bmath{\theta} \vert ^2/(2\sigma^2_w)) & \vert \bmath{\theta}\ \vert < 3\sigma_w, \\
    0 & \rm{otherwise}
  \end{cases}
\end{align}

\noindent where $\sigma_w$ is the aperture width. This significance is defined as the ratio between the aperture mass and its uncertainty, $M_{\rm{ap}}/\sigma_{M_{\rm{ap}}}$ \citep{Bartelmann01}, where

\begin{equation}
M_{ap} = \sum_{i} w_{\rm{g}}(\vert \bmath{\theta} - \bmath{\theta_i} \vert )g_{i,t}
\end{equation}

and

\begin{equation}
\sigma_{M_{\rm{ap}}}= \sqrt{\frac{1}{2}\sum_{i}^{\textcolor{white}{2}} w_{\rm{g}}^2 (\bmath{\theta} - \bmath{\theta_i}) \bmath{g_i}^2} \; ,
\end{equation}

with $g_{i,t}$ being the tangential reduced shear of the $i$th galaxy with respect to $\theta$. The width used for the aperture is $\sigma_w =$ 3 arcmin. Stacking the profile of our cluster sample we detect a significance peak of the aperture masses of $16 \sigma$ for the colour-magnitude selected sample. The significance map is nicely centred on the mean centre position of our cluster sample, as can be seen in Fig. \ref{fig:massmap}.

\subsection{Surface Mass Density Profile}
\label{dens_prof}

\ \ Stacking different clusters implies calculating the differential surface mass density profile $\Delta \Sigma(R)$ for every cluster-source pair rather than the pure shear signal $\gamma_{\rm{t}}$ since the first one explicitly takes the different angular diameter distance of the clusters into account. The differential surface mass density profile is defined as the difference between the mean surface mass density enclosed by radius R and the mean surface mass density at radius R,

\begin{equation}
\Delta \Sigma(R) = \overline{\Sigma} (<R) - \overline{\Sigma (R)},
\end{equation}

\noindent and is related to the tangential shear $\gamma_{\rm{t}}$ by:

\begin{equation}
\label{delsig}
\Delta \Sigma (r) = \Sigma_{\rm{cr}}   \gamma_t (r) 
\end{equation}

\noindent where the critical surface mass density is:

\begin{equation}
\label{scrit}
\Sigma_{\rm{cr}} = \frac{c^2}{4 \pi G} \frac{D_{\rm{s}}}{D_{\rm{d}} D_{\rm{ds}}},
\end{equation}

To compute $\Sigma_{\rm{cr}}$ we adopt the spectroscopic redshift of the BCG and the mean value of $\beta$ given by equation \ref{betacm}. 

For each radial bin we calculate the weighted average differential mass density profile using the lensing weight \emph{w} and the critical density $\Sigma_{\rm{cr}}$ \citep{Velander14, Hudson15}:

\begin{equation}
\label{eq:delta-sigma-data}
 \Delta \Sigma(r)  = \frac{\sum w_{\rm{ls},i}\gamma_{\rm{t},i} \Sigma_{\rm{cr},i}}{\sum w_{\rm{ls},i}}
\end{equation}

\noindent where the sum is over all lens-sources pairs. The minimum variance weight $w_{\rm{ls}} = w \Sigma_{\rm{cr}}^{-2}$ is sensitive both to the lensing geometry and the uncertainty in the ellipticity measurement: $\Sigma_{\rm{cr}}^{-2}$ down-weights lens-source pairs which are close in redshift while \emph{w} takes into account the ellipticity measurement error and the intrinsic shape noise.





In the innermost radial bins the assumption that we are probing exclusively weak shear signal may not be valid. In fact what we measure from galaxy shapes is the reduced shear \emph{g}, which relates to the true shear $\gamma$ by the relation $g = \gamma / (1 - \kappa)$, where $\kappa = \Sigma/ \Sigma_{\rm{cr}}$. When measuring the signal around a lens in the weak lensing regime ones assume that $g_{\rm{t}} \approx \gamma_t$, but at the smallest scale we are probing, this approximation fails. To take this effect into account we apply a correction term following \citet{Mandelbaum06}. The surface mass density profile corrected for the non-weak shear effect is calculated by:

\begin{equation}
\widehat{\Delta \Sigma} = \Delta \Sigma + \Delta \Sigma \ \Sigma \ \mathfrak{L}_z
\label{Lz}
\end{equation}

\noindent where $\mathfrak{L}_z = \left \langle \Sigma_{\rm{cr}}^{-2} \right \rangle / \left \langle \Sigma_{\rm{cr}}^{-1} \right \rangle$ is estimated averaging $\Sigma_{\rm{cr}}$ over the values of each lens-source pair in the stacked sample.  

The properties of our sample are derived by fitting the observed differential surface mass density profile to a theoretical model for dark matter halos. We make use of the halo model prescription \citep{Seljak00, Cooray02} and the off-centring formalism \citep{Johnston07} to include the different contributions to the final observed lensing signal through the equation:

\begin{equation}
\label{h_model}
\Delta \Sigma (R) = S_M [p_{\rm{cc}} \Delta \Sigma_{\rm{NFW}} (R)  + (1 - p_{\rm{cc}}) \Delta \Sigma^{\rm{off}}_{\rm{NFW}} (R) + \Delta \Sigma_{2h}]
\end{equation}

where we aim to constrain the parameters related to the NFW profile $M_{200}$, $c_{200}$. The nuisance parameters are $p_{\rm{cc}}$, the fraction of clusters correctly centred, $\sigma_{\rm{off}}$, the off-centring scale which enters in the off-centred profile $\Sigma^{\rm{off}}_{\rm{NFW}}$ and $S_m$, the parameter accounting for measurement systematic uncertainties (as defined in eq. \ref{Sm}).
We describe each component in detail in the next sections.

\subsubsection{Main Halo contribution}
\label{nfw_prof}

\ \ Previous results from numerical simulations and observations have found that the universal NFW profile is a good description of dark matter halos (e.g. \citealt{Bullock01, Wang09, Okabe13, Umetsu14}). Other studies have shown that dark matter halos can deviate from the NFW profile \citep{Navarro04, Gao08, Dutton14, Klypin14}, being better described by the Einasto profile \citep{Einasto65}. The Einasto profile has an extra free parameter $\alpha$ related to the shape of the profile in addition to the same two structural parameters describing the NFW profile. For the present work we decided to adopt the NFW profile when modelling the main halo (the stacked signal of the galaxy clusters) as: it is a simpler model that provides a good fit to the data and; it also makes the comparison with other studies easier as different observational analysis and theoretical prediction are based on the same profile.

The NFW density profile is described by

\begin{equation}
\rho(r) = \frac{\delta_{\rm{c}} \rho_{\rm{crit}}(z)}{(r/r_s)(1 + r/r_s)^2}
\end{equation}

\noindent with $\delta_{\rm{c}}$ being the characteristic over-density of the halo,

\begin{equation}
\delta_c = \frac{200}{3} \frac{c^3}{\rm{ln}(1 + c) - c/(1 + c)}
\end{equation}

\noindent $\rho_{\rm{crit}}(z)$ the critical density of the universe at redshift z,

\begin{equation}
\rho_{\rm{crit}} = \frac{3 H^2(z)}{8 \pi G}
\end{equation}
 
\noindent where H(z) is Hubble's parameter at the given redshift and G is gravitational constant, and $r_s$ the scale radius. $r_s$ can be defined in terms of the radius $r_{\Delta}$, inside of which the main density is $\Delta$ times the critical density of the universe, and the concentration parameter of the halo, $r_s = r_{\Delta}/c_{\Delta}$. As the mass enclosed in a given radius can be inferred through the relation  

\begin{equation}
M_{\Delta} = \frac{4 \Delta \pi \rho_{\rm{crit}}(z) r_{\Delta}^3}{3} 
\end{equation}

\noindent the NFW profile is fully specified by the mass $M_{\Delta}$ and the concentration $c_{\Delta}$ of the halo.

We follow the formalism presented in \citet{Wright00} when describing the main halo component, making use of the analytical expressions for the NFW profile when fitting $\Delta \Sigma$ to the data.

\subsubsection{The two-halo term}
\label{2halo}

\ \ The NFW profile appears to be a good description of the halo density profile when we look at a radial range going from the inner regions out to approximately the virial radius. But on larger scales the profile begins to be dominated by the surrounding matter concentrations around the main halo, as structures tend to cluster in the Universe. 

We describe the contribution from the neighbouring halos in a similar way to that presented in \citet{Johnston07}, including a two-halo term (e.g. \citealt{Seljak00, Mandelbaum05}) in our model of the density profile:

\begin{equation}
\rho_{2h} = b(M_{200}, z)\ \bar{\rho}_m(z)\ \xi_l (r, z)
\end{equation}

\noindent where $b(M_{200}, z)$ is the linear bias parameter for dark matter halos \citep{Sheth99, Seljak04}, $\xi_l (r, z)$ is the linear matter auto-correlation function at redshift z and $\bar{\rho}_m(z) = \Omega_m \rho_{c,0} (1+z)^3$ is the mean density at redshift z. For the bias parameter we adopt the relation proposed in \citep{Tinker10}

The function $\xi_l (r, z)$ is given in terms of the growth function $D(z) = \frac{D_+(a)}{D_+(1)}$, with

\[
    D_+(a) = H(a) \int_0^a \frac{1}{(a' H(a'))^3} da' ,
\]

\noindent $\sigma_8$ and the linear correlation function $\xi_l (r)$ at redshift $z = 0$ computed from the linear power spectrum derived from \emph{CAMB} \citep{camb} for the \emph{Planck 2015} cosmology \citep{Planck15}: 

\[
    \xi_l (r, z) = D(z)^2\ \sigma_8^2\ \xi_l (r)
\]

The final contribution of the two-halo term to the lensing signal is expressed by the differential surface density profile $\Delta \Sigma_{2h} (R)$. If we define an effective bias parameter $B(M_{200}, z)$: 

\begin{equation}
B(M_{200}, z) = b(M_{200}, z)\ \Omega_m\ \sigma_8^2\ D(z)^2\
\end{equation}

\noindent we write the final contribution as:

\begin{equation}
    \Delta \Sigma_{2h}(R) = B(M_{200}, z)\ \Delta \Sigma_l(R)
\end{equation}

\noindent where the term $\Delta \Sigma_l (R)$ is again described as the difference between the mean density inside a radius \emph{R} and the mean density on the radius \emph{R}, $\overline{\Sigma_l} (<R) - \overline{\Sigma_l (R)}$, with

\begin{equation}
    \Sigma_l (R) = (1 + z)^3 \rho_{c,0} \int dy\ \xi_l((1+z) \sqrt{y^2 + R^2})
\end{equation}


\subsubsection{Off-centring correction}
\label{off}

\ \ The measured lensing signal of an object is highly dependent on the chosen halo centre. As a consequence, in the presence of a shift between the chosen centre and the weak lensing peak, the inferred lens mass and the concentration parameter will be underestimated. It is usually assumed that the BCG of a cluster corresponds to the centre of the halo, but this assumption may not be entirely justified, since cluster finder algorithms may misidentify the real BCG and even if they do identify correctly, there is evidence that the BCGs often do not correspond to the mass centre \citep{Johnston07, Oguri10, Zitrin12}. Some works have shown that the central galaxy can be moving through the potential well of the halo (e.g. \citealt{Bosch05, Gao06, Johnston07, Hikage13}).

When modelling a stack of individual clusters one must take into account the fraction of correctly centred halos and the fraction of miscentred halos in the sample. We adopt a similar approach as the one described in \citet{Johnston07, Ford15}, modelling both the fraction of correctly centred halos and the width of the off-set distribution. The adopted distribution is a one-dimensional cut through a two-dimensional Gaussian:

\begin{equation}
P(R_{\rm{off}}) = \frac{R_{\rm{off}}}{\sigma_{\rm{off}}^2}\ \rm{exp} \left [ -\frac{1}{2} \left(\frac{R_{\rm{off}}}{\sigma_{\rm{off}}}\right)^2 \right]
\end{equation}

This offset distribution has the effect of turning the original $\Sigma(R)$ profile, approximately flat in the inner region, into a smoothed radial profile showing a decrease in small scales, given by

\begin{equation}
\Sigma^{\rm{off}}(R) = \int_{0}^{\infty} \Sigma(R\mid R_{\rm{off}}) P(R_{\rm{off}})\ dR_{\rm{off}}
\end{equation}

\noindent with

\begin{multline}
 \Sigma(R\mid R_{\rm{off}}) = \frac{1}{2\pi} \int_{0}^{2\pi} \Sigma(r)\ d\theta, \\  r = \sqrt{R^2+R_{\rm{off}}^2 - 2RR_{\rm{off}} cos(\theta)}   
\end{multline}

\noindent and $\theta$ the azimuthal angle. Following the main halo contribution we adopt the NFW analytical expression to write $\Sigma(r)$. The smoothed differential surface mass density profile is then given by $\Delta \Sigma^{\rm{off}}_{\rm{NFW}} = \overline{\Sigma^{\rm{off}}}(<R) - \Sigma^{\rm{off}}(R)$, where

\begin{equation}
    \overline{\Sigma^{\rm{off}}}(<R) = \frac{2}{R^2} \int_0^R \Sigma^{\rm{off}}(R') R' dR'  
\end{equation}

\subsection{The covariance matrix}
\label{cov_ma}

\ \ A comparison between observational data and theoretical model must consider uncertainties and their correlations between the radial bins, invoking the use of the covariance matrix when fitting the data.

This covariance matrix has contributions coming from shape noise, uncorrelated large-scale structure (LSS) along the line of sight \citep{Hoekstra03, Hoekstra11, Umetsu14} and instrinsic variations of cluster profiles at fixed mass \citep{Gruen11, Gruen15}. 

We estimated the uncertainty covariance matrix by resampling the 27 individual cluster catalogues through 10000 sets of bootstrap iterations, and calculating the matrix as

\begin{multline}
\label{cov}
C(i, j) = \left [  \frac{N}{N - 1} \right ]^2 \frac{1}{N} \sum_{k=1}^N \left [ \Delta \Sigma_k (R_i) - \overline{\Delta \Sigma}(R_i) \right ] \times \\
\left [ \Delta \Sigma_k (R_j) - \overline{\Delta \Sigma}(R_j) \right ]
\end{multline}

\noindent where N is the number of bootstrap sets, R is the radial bin, $\overline{\Delta} \Sigma(R)$ is the average differential surface mass density over all the bootstrap realisations at a specific radial bin and $i$ and $j$ are indexes to radial bins. The error bars on figure \ref{full} are given by the square-root of the diagonal elements of the covariance matrix. By bootstraping the sample of individual clusters we are taking into account the different contributions to the covariance matrix (shape noise, LSS and cluster profile variations). 

Although $C(i, j)$ is an unbiased estimator for the true covariance matrix it is noisy, so its inverse $C^{-1}(i, j)$ (the quantity used to fit the model) is not unbiased \citep{Hartlap07}. If $N_s$ is the number of independent samples, $N_D$ is the size of the data vector and the condition $N_s > N_D + 2$ is satisfied, we can apply a correction factor to get an unbiased estimator of $C^{-1}(i, j)$ \citep{Taylor13}:

\begin{equation}
\label{inv_cov}
    \widehat{C}^{-1}(i, j) = \frac{N_s - N_D -2}{N_s - 1} C^{-1}(i, j)
\end{equation}

We calculate $\widehat{C}^{-1}(i, j)$ by assuming $N_s$ to be the number of bootstrap realisations and $N_D$ the number of individual cluster catalogues.

\subsection{Profile Fitting}
\label{fit}

\ \ Our analysis aims to constrain the mean density profile parameters $M_{200}$ and $c_{200}$ of the stacked clusters sample. We adopt a Bayesian analysis and explore the parameters space with a Monte Carlo Markov Chain (MCMC) algorithm (\emph{emcee}, \citealt{emcee13}), where we fit the observed stacked $\Delta \Sigma$ profile to the halo model (section \ref{dens_prof}). The analysis is based in a background population selected by the colour-magnitude decision tree.  Results for the 5-bands photometric redshift point estimate are quoted in Appendix \ref{ap:resul}.

The posterior of our analysis can be written as

\begin{multline}
\mathcal{P}(M_{200},c_{200},\sigma_{\rm{off}},p_{\rm{cc}},S_m | r_i,\Delta \Sigma(r_i)) \propto  \\
\mathcal{L}(r_i,\Delta \Sigma(r_i)|M_{200},c_{200},\sigma_{\rm{off}},p_{\rm{cc}},S_m) \times \\
\Pi(M_{200}) \times  \Pi(c_{200}) \times \Pi(\sigma_{\rm{off}}) \times \Pi(p_{\rm{cc}}) \times \Pi(S_m)
\end{multline}

\noindent where $\mathcal{L}$ is the likelihood of the data given the model and $\Pi$ are the priors on each model parameter.

The (log-)likelihood, assuming Normal uncertainties, is defined as  
\begin{multline}
ln \mathcal{L} = -\frac{1}{2} \sum_{i,j} (\Delta \Sigma_{\rm{data},i} - \Delta  \Sigma_{\rm{model},i})^T \times \widehat{C}^{-1}_{i,j} \times \\
(\Delta \Sigma_{\rm{data},j} - \Delta \Sigma_{\rm{model},j})    
\end{multline}

\noindent whith $\Delta \Sigma_{\rm{data}}$ as the differential surface mass density measured in the $i$,$j$-th radial bins (eq. \ref{eq:delta-sigma-data}), $\Delta \Sigma_{\rm{model}}$ the model prediction for the surface mass density in the $i$,$j$-th radial bin (Eq. \ref{h_model}) and $\widehat{C}^{-1}$ the inverse of the covariance matrix (Eq. \ref{inv_cov}). 

The multiplicative factor $S_m$ was defined in the end of section \ref{errors} and takes into account the systematic uncertainties arising from shape measurements and distance estimates, with a prior modelled as a normal distribution.  

Priors on $M_{200}$ and $c_{200}$ are assumed to be flat in the range $10^{13} < M_{200}/M_{\odot} < 10^{16}$, $0.1 < c_{200} < 20$, and zero otherwise. 

For the off-centring parameters we adopt a more informative prior described in what follows, in particular because there is an important degeneracy between these nuisance parameters with $M_{200}$ and $c_{200}$ which we want to avoid. 

\citet{Rykoff16} derived the off-centring parameters for the Dark Energy Survey - Science Verification sample by comparing the cluster centres given by redMaPPer to the centres based on South Pole Telescope SZ data, Chandra and XMM X-ray data. The constraints were derived for each sample individually and also through a joint likelihood using the three samples, and the results are consistent within the errors. They found the best-fit values, derived from the joint analysis, to be $p_{\rm{cc}} = 0.78^{+0.11}_{-0.11}$ and $\sigma_{\rm{off}} = 0.32^{+0.08}_{-0.06} R_{\lambda}$, where $R_{\lambda}$ is a cluster radius given by redMaPPer. The constrained fraction of clusters well centred, $78^{+11}_{-11}\%$, is consistent with the prediction of $82\%$ given by redMaPPer. The redMaPPer predicted fraction is the mean value of the centring probability assigned to each clusters by the algorithm.

As our centres are assigned by the redMaPPer algorithm, we decide to adopt these empirical results in our analysis, considering that the off-set distribution is mostly based on choosing the wrong central galaxy - and that the bright central galaxies are detected both in SDSS and in deeper data. We write the priors of the off-centring parameters as a normal distribution for $p_{\rm{cc}}$ with mean $\mu = 0.78$ and width $\sigma = 0.11$, and as a log-normal distribution for $\sigma_{\rm{off}}$ with $\rm{ln}(\mu) = -1.13$ and $\sigma = 0.22$. Note that we have omitted the factor $R_{\lambda}$ in the $\sigma_{\rm{off}}$ parameter, as it spans a range of $0.9 - 1.1\ h^{-1}\ Mpc$ for our sample and the width of the prior is larger than this range. As a reference we quote the predicted fraction of correctly centred clusters given by redMaPPer for our sample, $\left \langle P_{\rm{cen}} \right \rangle = 0.82$. 

For the MCMC implementation we run 256 chains, each one containing 10000 steps. After testing for stability, running different numbers of chains with different sizes and burn-in regions and looking at the trace of each parameter, we checked that the final result is robust and that discarding the first 200 steps of each chain is enough to eliminate the burn-in part.

We test our method using mock catalogues of cluster-size halos generated from an N-body simulation for a flat $\Lambda$CDM cosmology \citep{Becker11}. We stacked the simulated halos in bins of $0.1 \times 10^{14} M_{\odot}$ for low z halos ($z = 0.25$) and of $1 \times 10^{14} M_{\odot}$ for the high z halos ($z = 0.50$), starting at $M_{200c} = 1 \times 10^{14} h^{-1} M_{\odot}$. We then run our Bayesian code fitting the shear profile from the simulation in the radial range $0.2 < R [Mpc] < 4$. For this exercise we use the halo model with two components: the main halo following an NFW profile plus the two-halo term, described in Sections \ref{nfw_prof} and \ref{2halo}. We let $M_{200c}$ and $c_{200c}$ as free parameters with a Gaussian prior for the concentration, first assuming the \citet{Duffy08} c-M relation and in a second run the \citet{Dutton14} relation, and an uninformative prior for the mass. The results on $M_{200c}$ are biased low by $\sim 5 - 8 \%$, in agreement with what was found in \citet{Becker11}, and are insensitive to the choice on the concentration prior. Since the purpose of this work is determining the best-fit NFW parameters (in terms of mass and concentration) for our cluster sample, we do not apply this as a bias correction to our fit results.

\begin{figure*}
\centering
\includegraphics[width=10cm]{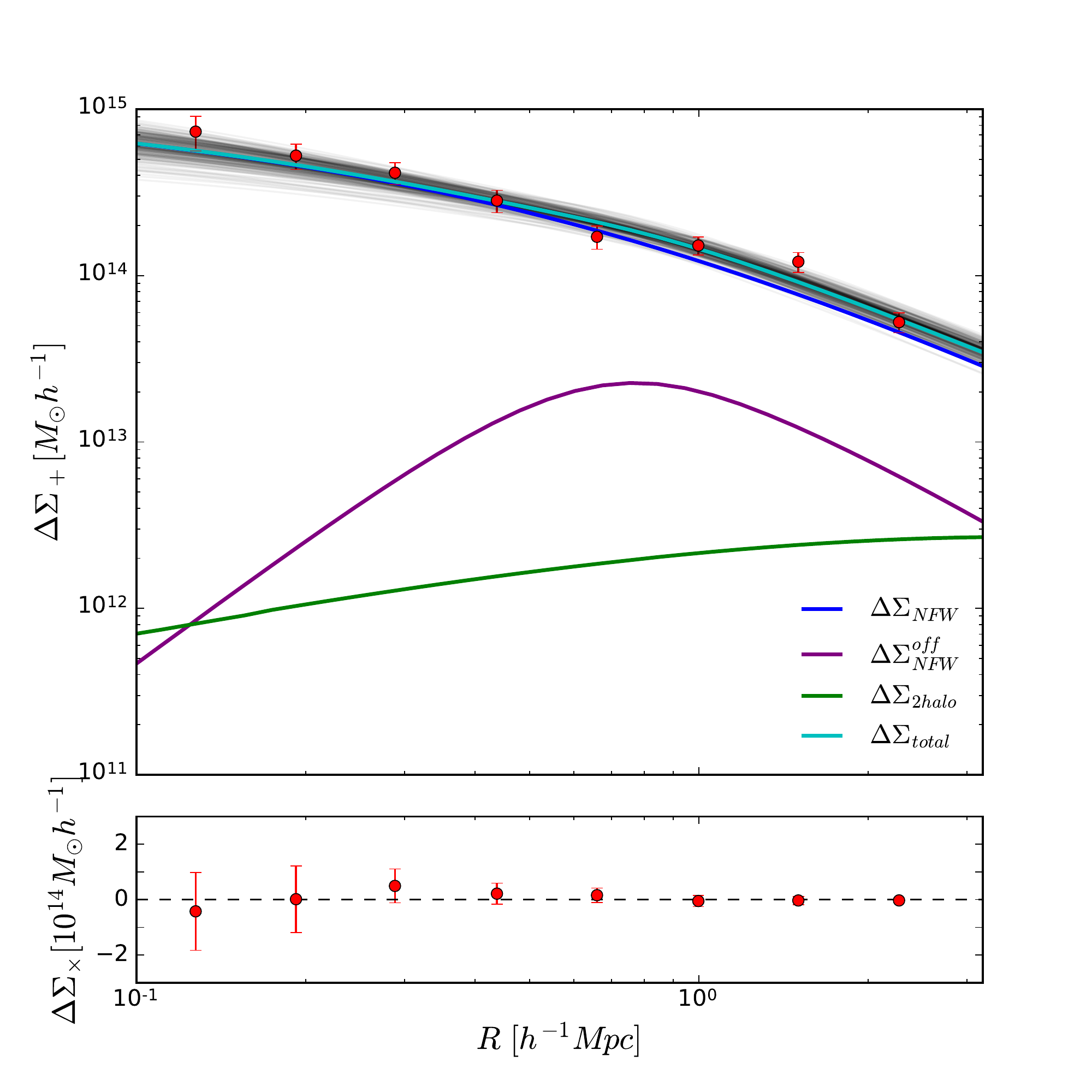}
 \caption{Stacked tangential shear profile of the 27 clusters in units of projected mass density, with 100 randomly selected MCMC realisations to sample the $\Delta \Sigma_{total}$ variance (grey lines). Each component of the halo model, as well as the total signal, is computed for the best-fit values and plotted separately. The lower panel shows the cross component of the shear signal, used to check for systematics.}
\label{full}
\end{figure*}

\subsection{Results}
\label{results}
  
\ \ Figure \ref{full} shows the stacked lensing signal, the overall fit and each component of the halo model (the main term described as an NFW profile, the two-halo term and the smoothed NFW profile corresponding to the off-centring component). The red dots are the measured differential surface mass density profile, with the error bars given by the square root of the main diagonal of the covariance matrix (equation \ref{cov}). We plotted 100 random MCMC realisation out of the total 256 chains, corresponding to the grey lines, to exemplify the parameter space exploration of the fitting. The different lines show each individual component of the halo model and the blue line shows the full model (described in \ref{dens_prof}) for the best-fit values extracted from the Bayesian analysis. 

To check for systematics in the weak lensing measurements we analysed the cross-component shear signal, $\Delta \Sigma_{\times}$. This signal corresponds to a rotation of 45 degrees with respect to the tangential shear $\Delta \Sigma_+$ and for an unbiased measurement of ellipticities we expect to get $\Delta \Sigma_{\times}$ compatible with zero. We can see in the lower panel of figure \ref{full} that in all bins our $\Delta \Sigma_{\times}$ signal is consistent with zero within $2\sigma$, indicating that there is no significant systematic in our observed shear signal.

The signal is detected with a $S/N \simeq 14$, calculated using the full covariance matrix $C_{i,j}$ \citep{Okabe15}:

\begin{equation}
\label{sn_matrix}
(S/N)^2  = \sum_{i,j} \Delta \Sigma_i\ C^{-1}_{i,j}\ \Delta \Sigma_j
\end{equation}

We find that the non-weak shear correction term $\Delta \Sigma\ \Sigma \mathfrak{L}_z$ (eq. \ref{Lz}) has the biggest impact in the innermost bin, changing the surface mass density profile by about $20\%$. The factor $\mathfrak{L}_z$ is $2.46 \times 10^{-16} h^{-1} Mpc^2 / M_{\odot}$ for our data.

\begin{figure*}
\centering
 \includegraphics[width=12cm]{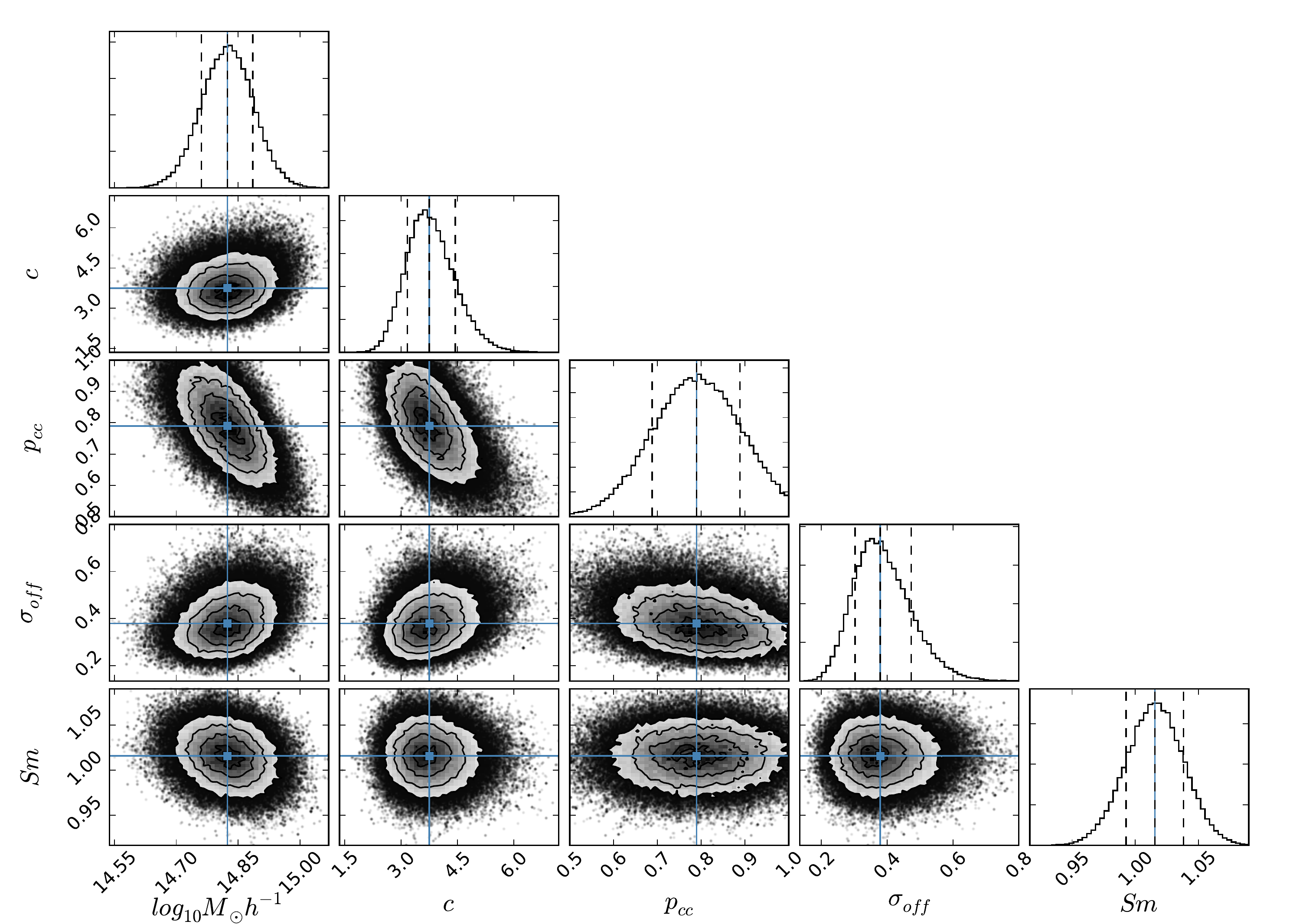}
 \caption{The one and two dimensional projections of the posterior distributions for the stacked profile parameters M$_{200c}$, c$_{200c}$ and the nuisance parameters related to the off-centring effect and systematic uncertainties. The blue dots represent the best-fit values quoted as the median of the distributions and the dashed vertical lines show the $68\%$ interval.}
\label{mcp}
\end{figure*}

The marginalised and the two-dimensional posterior distributions of the free parameters $M_{200}$, $c_{200}$ and the nuisance parameters $p_{\rm{cc}}$, $\sigma_{\rm{off}}$, $S_m$ are shown in figure \ref{mcp}. The posterior distribution of the off-centring parameter $p_{\rm{cc}}$ follows the adopted prior: both present mean values of 0.78 and variance of 0.1. The parameter $\sigma_{\rm{off}}$ is affected by the data but we still find a best-fit value in agreement with the mean of the prior within $1\sigma$: the log-normal prior has mean value of $0.32^{+0.08}_{-0.06}$ while the posterior distribution has $0.38^{+0.09}_{-0.08}$. The density profile parameters $M_{200}$, $c_{200}$ are well constrained by our data. The extra free parameter $S_m$ is marginalised and accounts for systematic uncertainties on shape and distance measurements, as described in the end of section \ref{errors}. 

In table \ref{tab:fit} we quote the median of the MCMC chains per parameter and the interval containing 68$\%$ of the points from the posterior probability distribution.
We also quote the mode of the 1D marginalised distributions and the values given by approximating them as a Gaussian and taking the mean and standard deviation - we see from figure \ref{mcp} that a Gaussian distribution is a good approximation for marginalised distributions of mass and concentration.

\begin{table*}
\centering
\caption{Summary of the posterior distributions.}
\begin{tabular}{|p{4.5cm}| p{2cm} p{2cm} p{2cm} p{2cm}|}
\hline
Adopted value & M$_{200c}$ $[10^{14} h^{-1} M_{\odot}]$ & c$_{200c}$ & $p_{\rm{cc}}$ & $\sigma_{\rm{off}}$ $[h^{-1} Mpc]$ \\
\hline
Median and 68\% interval & $6.6^{+1.0}_{-0.8}$ & $3.7^{+0.7}_{-0.6}$ & $0.8^{+0.1}_{-0.1}$ & $0.38^{+0.09}_{-0.08}$ \\
Mode of the 1D distribution & $6.6$ & $3.6$ & $0.8$ & $0.36$ \\
Gaussian approximation & $6.6 \pm 1.1$ & $3.8 \pm 0.6$ & $0.79 \pm 0.09$ & $0.38 \pm 0.08$ \\
\hline
\end{tabular}
\label{tab:fit}
\end{table*}

The 2D posterior distributions evidence the degeneracy between the off-centring parameters $p_{\rm{cc}}$, $\sigma_{\rm{off}}$ and the density profile parameters $M_{200}$, $c_{200}$. We calculated the correlation coefficient $\rho$ between $M_{200}$ and $p_{\rm{cc}}$ ($\rho_{\rm{M,pcc}}$), $M_{200}$ and $\sigma_{\rm{off}}$ ($\rho_{\rm{M,soff}}$), $c_{200}$ and $p_{\rm{cc}}$ ($\rho_{\rm{c,pcc}}$), $c_{200}$ and $\sigma_{\rm{off}}$ ($\rho_{\rm{c,soff}}$) using the covariance and standard deviation extracted from the trace of each parameter, finding $\rho_{\rm{M,pcc}} = - 0.60$, $\rho_{\rm{M,soff}} = 0.27$, $\rho_{\rm{c,pcc}} = -0.56$ and $\rho_{\rm{c,soff}} = 0.20$. 
Both the 2D posterior distributions and the correlation coefficients support a strong correlation of $M_{200}$, $c_{200}$ with respect to $p_{\rm{cc}}$, while the correlation with $\sigma_{\rm{off}}$ is weaker. 

The parameters $c_{200}$ and $p_{\rm{cc}}$ influence the very central region of the density profile, and are therefore degenerate. Mass impacts the profile on all scales and presents the most important degeneracy with $p_{\rm{cc}}$, which can be explained by the influence of $p_{\rm{cc}}$ also on intermediate scales, up to a radius of $\sim 0.7$ Mpc (as can be seen from the off-centred component represented by the magenta line in figure \ref{full}). 

Finally we note that constraints obtained by the two background selection methods, the colour-magnitude decision tree and the photometric redshift point estimator, are in good agreement (see Appendix \ref{ap:resul} for the results based on the photo-z point estimate analysis and a comparison between the $\Delta \Sigma_+$ profiles). The difference between the best-fit values of $M_{200c}$ and $c_{200}$ is negligible when compared to the total error, demonstrating that the results are valid within the two methods. The agreement between the two selection criteria is also a strong evidence that the residual noise bias in the self-calibrating \emph{lensfit} is not affected by the photometric redshift or colour-magnitude cuts.

\subsubsection{Impact of the off-centring priors}

\ \ Given the degeneracy between $M_{200}$, $c_{200}$ and the off-centring parameters we explore the results derived when adopting different priors for $p_{\rm{cc}}$ and $\sigma_{\rm{off}}$ as well as for fixed values to check the impact on  $M_{200}$ and $c_{200}$. We first fix the $p_{\rm{cc}}$ value to be equal the expectation value given by redMaPPer for our sample, $\left \langle P_{cen} \right \rangle = 0.82$, and $\sigma_{\rm{off}} = 0.42 h^{-1} Mpc$, where the choice of this value comes from the literature \citep{Johnston07, Oguri10}, letting the parameters $M_{200}$ and $c_{200}$ be constrained by our data. We then use the fixed value of $p_{\rm{cc}} = \left \langle P_{cen} \right \rangle = 0.82$ and let $\sigma_{\rm{off}}$ be constrained by the data, adopting a flat prior in the range 0.01 - 0.8. Finally we set both $p_{\rm{cc}}$ and $\sigma_{\rm{off}}$ as free parameters with a flat prior in the range of 0.5 - 1 and 0.01 - 0.8, respectively. Table \ref{tab:pcc} summarises these results, given by the 16th, 50th and 84th percentiles of the samples.

\begin{table*}
\centering
\caption{Best-fit values for different choices on the off-centring parameters.}
\begin{tabular}{|p{3.5cm}| p{3.5cm} p{2cm} p{2cm} p{2cm} p{2cm}|}
\hline
Prior on $p_{\rm{cc}}$ & Prior on $\sigma_{\rm{off}}$ & M$_{200c}$ $[10^{14} h^{-1} M_{\odot}]$ & c$_{200c}$ & $p_{\rm{cc}}$ & $\sigma_{\rm{off}}$ \ \ \ \ $[h^{-1} Mpc]$ \\
\hline
Fixed value & Fixed value & $6.5^{+0.7}_{-0.7}$ & $3.7^{+0.6}_{-0.5}$ & 0.82 & 0.42 \\
Fixed value & Flat, $0.01 < \sigma_{\rm{off}} < 0.8$ & $6.6^{+0.8}_{-0.7}$ & $3.7^{+0.6}_{-0.5}$ & 0.82 & $0.6^{+0.2}_{-0.2}$ \\
Flat, $0.5 < p_{\rm{cc}} < 1$ & Flat, $0.01 < \sigma_{\rm{off}} < 0.8$ & $7.6^{+1.5}_{-1.6}$ & $4.4^{+0.9}_{-0.9}$ & $0.66^{+0.20}_{-0.12}$ & $0.63^{+0.11}_{-0.17}$\\
\hline
\end{tabular}
\label{tab:pcc}
\end{table*}

We see that the best-fit values for mass and concentration for different approaches on the off-centring formalism agree within $1\sigma$ with the results constrained by the full model adopting the priors on $p_{\rm{cc}}$ and $\sigma_{\rm{off}}$ from \citet{Rykoff16}. When fixing $p_{\rm{cc}}$ and $\sigma_{\rm{off}}$ and for a fixed $p_{\rm{cc}}$ with $\sigma_{\rm{off}}$ as a free parameter we recover the constraints on $M_{200}$ and $c_{200}$ in good agreement with the values derived by the full model, while the errors decreases by $\simeq 15 - 20\%$. For flat priors on $p_{\rm{cc}}$ and $\sigma_{\rm{off}}$ we recover a lower fraction of correctly centred clusters and a higher off-set scale, implying in higher values of mass and concentration for the same sample with errors significantly increased by $\sim 50 - 100\%$.

We note that for $p_{\rm{cc}} = \left \langle P_{\rm{cen}} \right \rangle = 0.82$ and $\sigma_{\rm{off}}$ following a flat prior we recover a marginally larger off-set scale ($\sigma_{\rm{off}} = 0.57^{+0.16}_{-0.20} h^{-1} Mpc$ compared to $0.42 h^{-1} Mpc$ when fixing the parameter and to $\sigma_{\rm{off}} = 0.38^{+0.09}_{-0.08} h^{-1} Mpc$ for the full model), but this has a negligible impact on $M_{200}$ and $c_{200}$. This result together with the correlation seen from the 2D posterior distributions indicate that the main degeneracy is between the density profile parameters and $p_{\rm{cc}}$, with a weaker dependence on the off-set scale $\sigma_{\rm{off}}$: the suppression in the inner part of the profile is attributed to a large fraction of miscentred halos leading to a over-correction of the off-centring effect, which in turn allows for higher values of $M_{200}$ and $c_{200}$.

\subsection{Impact of the stacking procedure}
\label{simul_mc}

\ \ In this work we opt to do a single fit of a stacked profile instead of fitting individual clusters (with or without the use of relations that would decrease the number of free parameters). However, we want to compare our data with models of several individual clusters following a mass distribution and presenting a intrinsic scatter in concentration for a fixed mass. 
To check whether or not we introduce a bias in our results by stacking the clusters we perform simulations. We did 1000 iterations ($k = 1, ..., 1000$) of the following procedure:


\begin{enumerate}
\item Given the $i=1,..,27$ observed richness of the clusters ($\lambda_i$) we draw a cluster mass $M_i$ from the \citet{Simet16} mass-richness relation with 0.25 dex  intrinsic scatter.
\item From this mass $M_i$ and the individual cluster redshift $z_i$, we draw a concentration ($c_i$) using the \citet{Dutton14} relation with 0.1 dex intrinsic scatter.
\item With $M_i$, $c_i$ and $z_i$ we create an artificial differential surface mass density profile $\Delta\Sigma_i(r)$ for each cluster.
\item We stack the 27 individual $\Delta\Sigma$ profiles and fit it using the same covariance matrix derived earlier, thus obtaining a pair of M$_{fit}$ and c$_{fit}$ values. 
\item We compute the mean $M_k$, $c_k$ of the input parameters $M_i$, $c_i$, weighted by the lensing weight $w_{ls}$ (eq. \ref{eq:delta-sigma-data}). 
\end{enumerate}

We then compare by how much M$_{fit}$, c$_{fit}$ deviate from the weighted-mean of the true input parameters $M_k$, $c_k$ on average and in terms of \emph{rms}.

Averaging over the 1000 realisations, the offset between the natural logarithms of the true weighted-mean mass to the fitted mass, $ln(\overline{M_k}) - ln(\overline{M_{fit}})$, is $+0.016$. In terms of \emph{rms} the values deviate by $0.018$. The true weighted-mean concentration differs from the mean fitted concentration, $\overline{c_k} - \overline{c_{fit}}$, by $+0.12$, with \emph{rms} of $0.13$ over all realisations. Evaluating the concentration for the true weighted-mean mass at the mean redshift of the stacking ($\bar{z} = 0.50$) using the \citet{Dutton14} relation gives a value that differs from $\overline{c_k}$ by $-0.05$ and from $\overline{c_{fit}}$ by $+0.07$. 

From these simulations we found no evidence that the stacking procedure brings about a bias as the offsets are consistent with zero, given the scatter ($+0.016 \pm 0.018$ in mass, $+0.12 \pm 0.13$ in concentration). Comparing the \emph{rms} from these simulations to our $1\sigma$ confidence level from the Bayesian stacked analysis (Table \ref{tab:fit}) we found that the latter is about one order of magnitude larger for the mass and 5 times larger for the concentration then the former. We thus conclude that any bias introduced by the stacking procedure (less than $2\%$ in mass) should be negligible compared to the statistical plus systematic (from shape and redshift measurements) uncertainties ($\sim 15\%$ in mass). 

\subsection{Comparison with other studies}
\label{compare}

\ \ In this section we compare our sample and the results constrained by our data to other observational studies. We also analyse the performance of different numerical simulations in predicting our observed concentration. 

We found that the impact of selection effects in our c-M results are likely subdominant compared to the statistical uncertainties. Our Bayesian analysis results in errors of the order of $12-15\%$ in mass and $15-20\%$ in concentration. From simulations presented in the literature we know that: X-ray selection can cause a bias of $\sim 10\%$ in $c$ \citep{Meneghetti14}, caused by the large fraction of relaxed systems; an optically selected sample leads to overestimated richness due to orientation bias, and the WL masses estimated for a stacked optically selected sample has a positive bias of $3-6\%$ \citep{Dietrich14}, and likely a bias in $c$. The estimation of the factor that would arise for our sample selection is outside the scope of this work, but we assume that it would be smaller than what was found in \citet{Meneghetti14}, since the X-ray data we select on is more noisy and thus less constraining. We can expect that our richness selection could contribute to the final selection bias, but also only at the few percent level. From these numbers we assume that the possible shift in our c-M relation caused by selection effects is encompassed within our confidence levels, thus making the comparison between our results to other observational studies and numerical predictions valid.

\subsubsection{The comparison samples}

\ \ An interesting observational result to be compared with our work due to the similarity in the mass range but in a lower redshift interval is the study conducted by the LoCuSS collaboration \citet{Okabe15}: the samples contain high mass galaxy clusters at the given redshift interval ($\bar{z} = 0.50$ for the CODEX sample and $\bar{z} = 0.23$ for the LoCuSS). The LoCuSS mass-concentration relation was derived for a stacked sample of 50 individual X-ray selected clusters based on RASS catalogues, satisfying the criteria $0.15 \leqslant z \leqslant 0.3$ and L$_X/$E(Z) $> 4.1 \times 10^{44}\ \rm{erg}\ \rm{s}^{-1}$, with L$_X$ in the $0.1 - 2.4$ keV band. Because no other physical property of the clusters enters in the selection function besides L$_X$ the authors argued that the estimated c-M relation should be unbiased.

The CLASH collaboration \citep{Clash16} determined the mass-concentration relation based on strong and weak lensing and magnification measurements for a subsample of 16 X-ray regular and 4 high-magnification clusters taken from the CLASH sample described in \citet{Postman12}. This initial CLASH sample comprises 20 clusters with X-ray temperature $> 5 keV$ and showing a smooth X-ray morphology, with no lensing selection. From their X-ray regular sample we have selected the six higher redshift clusters, spanning a range of $0.391 < z < 0.696$, and the stacked density profile of the 16 X-ray regular clusters with $\bar{z} = 0.35$. For the last we have selected the model assuming a NFW description for the main halo and the large-scale structure contribution with the bias b$_h$(M) as a scaling factor. Numerical simulations for the CLASH sample have found that the X-ray selected sample is mainly composed by relaxed halos \citep{Meneghetti14}, and for the most regular halos the concentration is about $11\% \pm 3$ higher than the average over all simulated halos. A better agreement between the observed concentration-mass relation and numerical simulations is achieved when selection effects are taken into account.

The LoCuSS and CLASH studies applied conservative cuts to select background galaxies, controlling the contamination at the $\sim 2\%$ level. By controlling the contamination to the few percentage level they decided to not include a statistical correction factor that takes into account dilution effects, the so-called boost factor. This correction assumes that the background number density profile should be flat and neglect magnification effects that can cause enhancement or depletion of this profile \citep{Umetsu14, Okabe15}. We note that we also choose to apply a conservative cut and do not include the boost factor in our analysis.

Another similarity between the three samples is the adopted cluster centres corresponding to the BCG. 

\begin{figure}
\def\tabularxcolumn#1{m{#1}}
\begin{tabularx}{\linewidth}{@{}cXX@{}}
\begin{tabular}{cc}
\includegraphics[width=7.8cm]{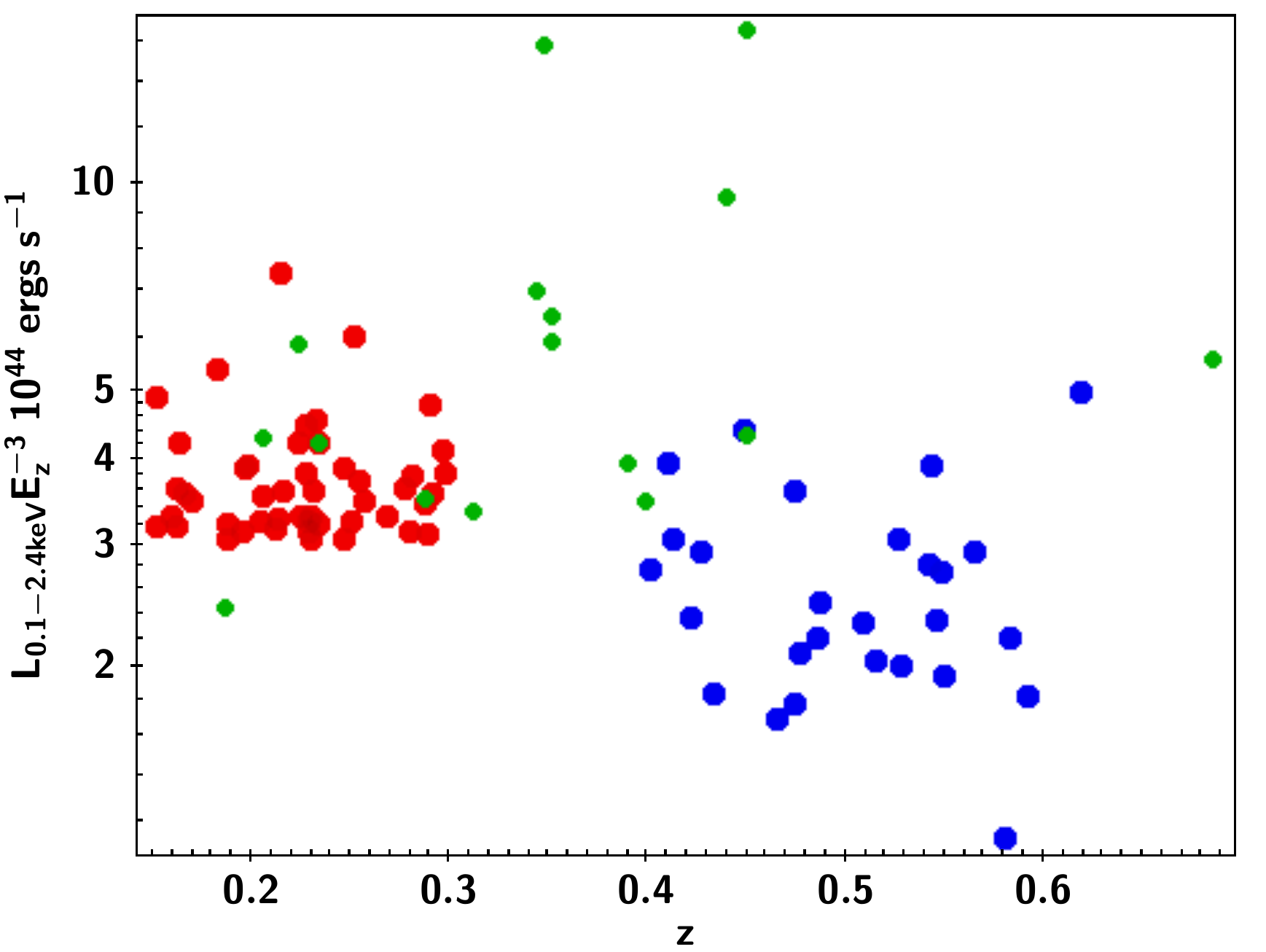} \\
\includegraphics[width=7.8cm]{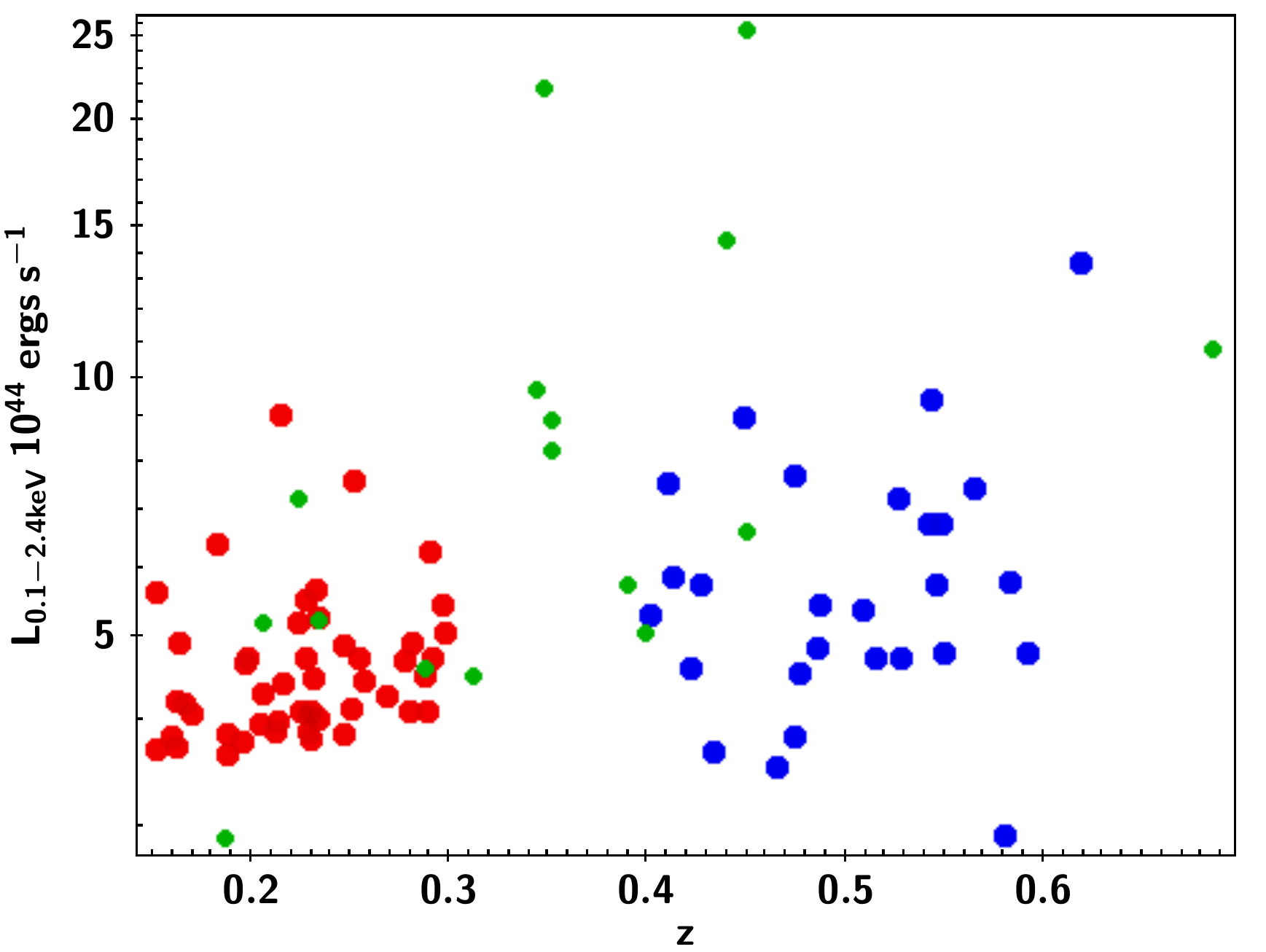}
\end{tabular}
\end{tabularx}
 \caption{Distribution of X-ray luminosity with redshift for the CODEX CFHT WL clusters (blue symbols), LoCuSS (red symbols) and CLASH (green symbols) samples. The top panel shows the distributions for L$_X$ corrected for the redshift evolution assuming self-similarity and the bottom panel shows the uncorrected L$_X$.}
\label{codexmEz}
\end{figure}

Figure \ref{codexmEz} shows the distributions of X-ray luminosity with redshift for our 27 CFHT WL clusters, for the LoCuSS and for the CLASH samples. The top panel shows the L$_X$ corrected for the redshift evolution by the factor $E(z) = H(z)/H_0$ while the lower panel shows the uncorrected L$_X$. Our sample has on average lower values of corrected L$_X$ but similar values of uncorrected L$_X$ compared to the LoCuSS sample while CLASH probes higher L$_X$ clusters. 

To produce the comparison plots we first match the LoCuSS and CLASH clusters position to the CODEX SDSS redMaPPer catalogue and use our estimate of L$_{X}$ when available; we found 36 matches for LoCuSS and only 5 for CLASH, as most of them come from the southern hemisphere. Notwithstanding we do not expect a substantial difference in the luminosity distribution caused by the inclusion of different sources, as L$_x$ estimates between different missions agree to within $10\%$, typically, e.g. \citet{Zhang04}. 

\begin{figure}
\includegraphics[width=8.2cm]{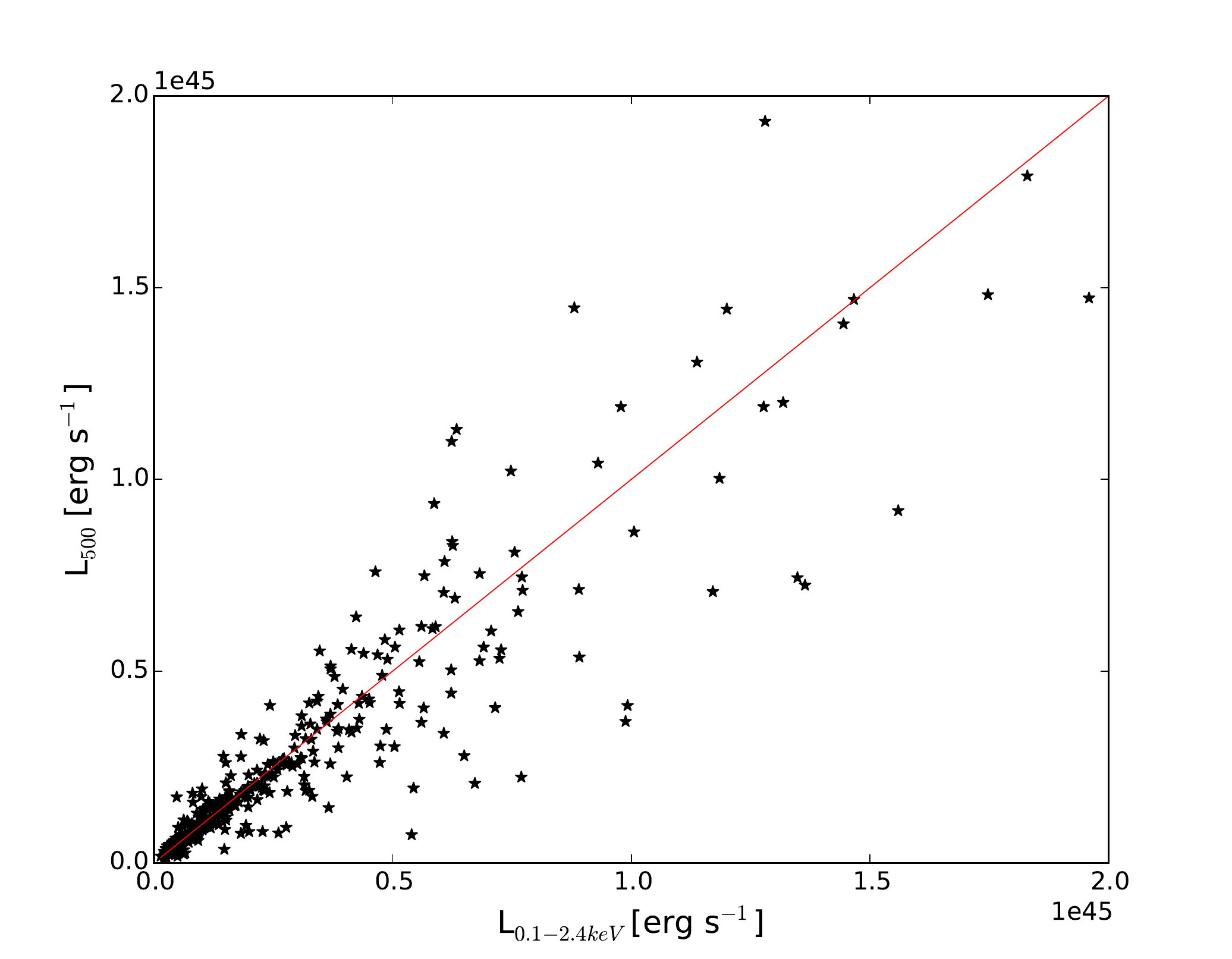} 
 \caption{Codex estimate of $L_{0.1-2.4keV}$ vs $L_{500}$ from the MCXC catalog. The red line indicates the one-to-one relation.}
\label{lx_mxcx}
\end{figure}

We also make use of the X-ray catalogue MCXC \citep{Piffaretti11}, which is based on publicly available ROSAT cluster catalogues, to check the homogeneity of the data. The MCXC catalogue has data homogenised to an over-density of 500 and comprises a sample large enough for statistical studies. The MCXC catalogue is part of the full CODEX sample. We match the positions and compare the $L_{500}$ X-ray luminosities presented in this catalogue with our estimate of $L_{0.1-2.4keV}$ in figure \ref{lx_mxcx}. The values are close to a one-to-one relation (red line) for most of the points, indicating that our luminosity estimates are homogenised as well and compatible with other luminosity estimates for ROSAT cluster catalogues. 

\begin{figure}
\centering
 \includegraphics[width=7.3cm]{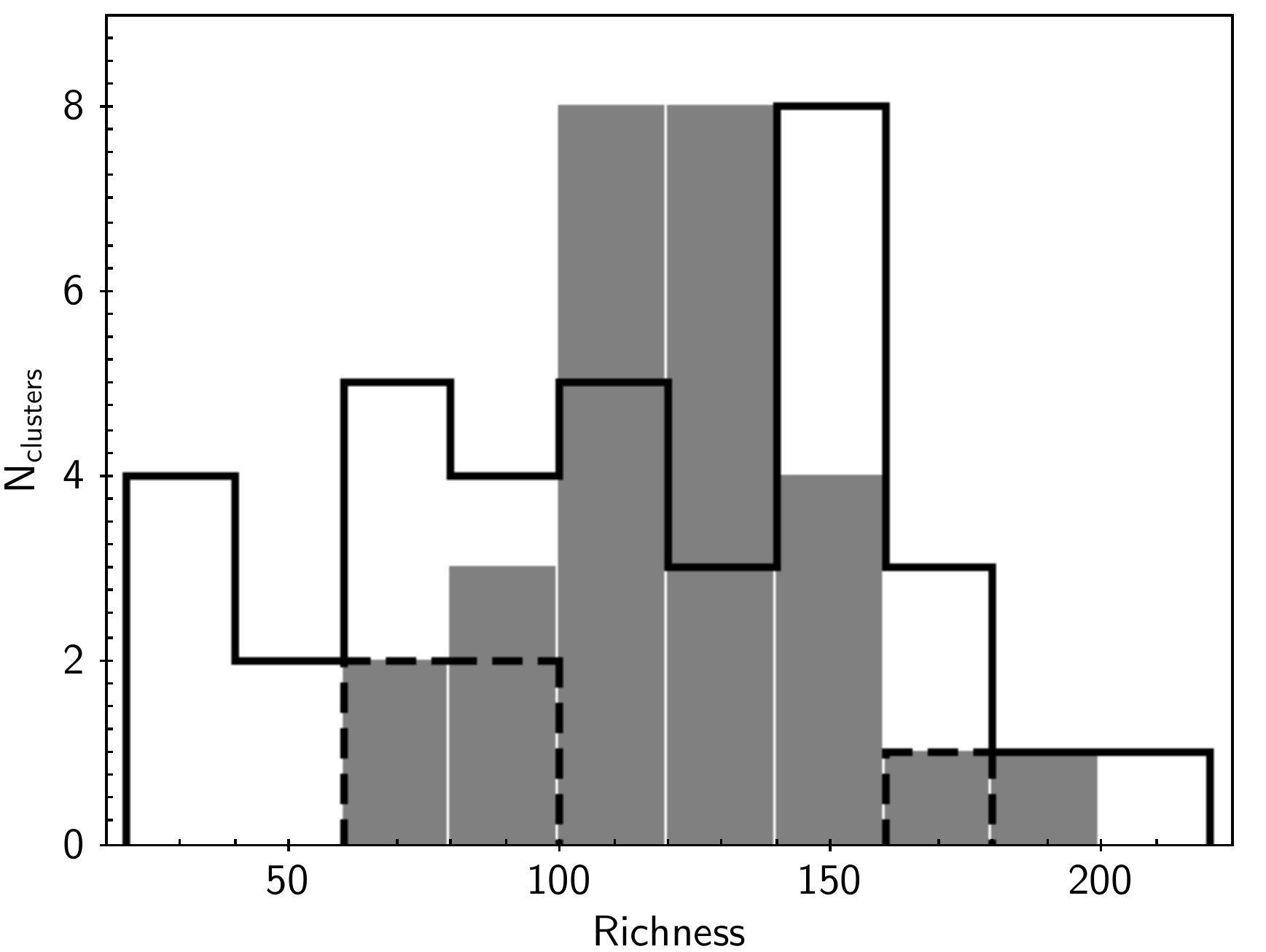}
 \caption{Richness distribution for the CODEX clusters (grey histogram) and for the matched LoCuSS (solid-line contour) and CLASH (dashed contour) clusters.}
\label{codexmz}
\end{figure}

We obtain a richness estimate for the comparative studies using the matched position of the LoCuSS and CLASH clusters to the CODEX catalogue. The richness distributions are shown in Figure \ref{codexmz}.. 

The richness distribution of our sample is similar to the LoCuSS distribution with the difference that we applied a cut for $\lambda > 60$. The LoCuSS sample has on average higher corrected L$_X$ but it contains a number of lower richness systems.

\begin{figure}
\centering
 \includegraphics[width=8.8cm]{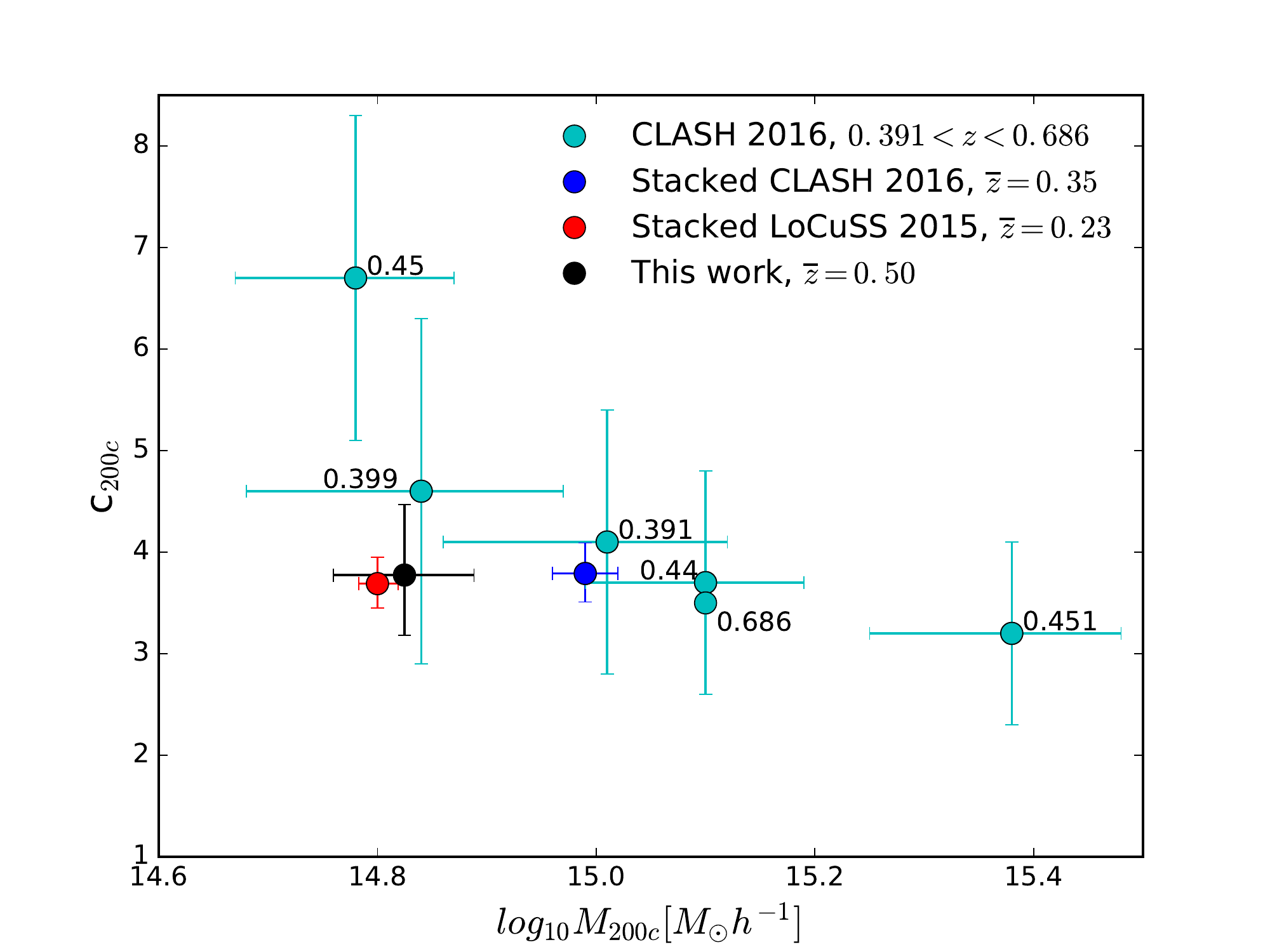}
 \caption{Best-fit c$_{200c}$ and M$_{200c}$ parameters for the CODEX mass-concentration relation at $\bar{z} = 0.50$ (black dot). The red dot is the relation found by \citealt{Okabe15} from the stacked sample of 50 clusters at $\bar{z} = 0.23$. The cyan dots are results for individual clusters with $z > 0.39$ (individual redshifts are displayed together with the individual measurements) and the blue dot is the best-fit value for the stacking of the 16 X-ray selected clusters (redshift range of $0.187 < z < 0.686$), both from \citealt{Clash16}.}
\label{complocuss}
\end{figure}

\subsubsection{Comparison between observed masses and concentrations}

\ \ In figure \ref{complocuss} we compare our best-fit values for the mass and concentration with the results from the LoCuSS stacked sample and with the high-z individual and stacked CLASH sample. These results show that the observed concentration for clusters in the mass range of $5 \times 10^{14} M_{\odot} < M_{200c} < 10^{15} M_{\odot}$ is around $c = 3.7$ with no clear dependence on redshift for the range probed by the samples ($0.2 < z < 0.6$). 

We obtain a comparable (marginally larger) value of mass with the LoCuSS result.
If a richness cut similar to the one adopted in this work is applied to the LoCuSS sample (Okabe N. 2016 private communication), meaning a stacking of clusters with $\lambda_{\rm{SDSS}} \geqslant 60$, the LoCuSS best-fit mass value is increased from $6.37^{+0.28}_{-0.27} 10^{14} h^{-1} M_{\odot}$ to $6.88^{+0.39}_{-0.37} 10^{14} h^{-1} M_{\odot}$. 
Higher mass values of the CLASH sample are consistent with the higher richness, with the caveat that we sample only a small fraction of the CLASH sample on richness measurements, and with the higher X-ray luminosities.

Regarding the concentration we could expect a sign of evolution with redshift: numerical simulations predict a weak dependence of the concentration with the halo redshift, decreasing for higher-z at a fixed mass. This evolution is not apparent in the comparison between our analysis and LoCuSS as both samples statistically show the same value of concentration, a result that favours very little or no evolution for the concentration between $z = 0.2$ and $z = 0.5$. 

There are some differences between the modelling adopted in this work with the one adopted by LoCuSS that are worth noting. In \citet{Okabe15} the shear profile modelling includes three mass components: a central point mass associated with the BCG, the main halo described by an NFW profile and the two-halo term. In our halo model we describe the signal as a result of three components: the main halo, characterised by a NFW profile, an off-centering component and the two-halo term. The differences in the two approaches may imply in differences in the concentration estimates: a misidentified centre suppresses the shear signal in the central region reflecting in a lower value of concentration whereas an over-correction of the off-centring effect will result in a higher concentration, and the inclusion of a central point mass can also lead to a lower concentration for a given observed profile \citep{Umetsu14}.

The CLASH stacked concentration value is very close to our result, while the mass is higher. Ignoring the weaker dependence of $c$ on $z$, the expected relation between concentration and mass predicts a decreasing $c$ with increasing $M$ - though some studies, e.g. \citealt{Prada12}, found an upturn in the relation. The selection of relaxed halos could change this relation for the CLASH sample, as argued in \citet{Meneghetti14}, leading to higher concentration values.

When comparing our result with the individual CLASH cluster measurements at a similar redshift range, our concentration value is compatible within $1\sigma$ with all their constraints but the lowest mass system at $z = 0.45$. While there is a trend of lower values of concentrations for higher masses when we look at the individual CLASH measurements, the inclusion of our constraint favours a flatter slope. 

The model adopted for the CLASH sample \citep{Clash16} does not take into account the uncertainty in the cluster centre, which is assumed to be the BCG. It was argued that the typical offset between the BCG and the X-ray peak is negligible if compared to the radial range for the mass measurments.

\subsubsection{Concentration-mass relation from simulations}

\ \ We now turn our attention to the comparison of our results with the predictions from numerical simulations. Studies based on simulations have derived different relations for predicting the concentration of a dark matter halo given a mass and a redshift. We choose five different relations based on N-body cosmological simulations for this comparison, briefly described below. 

\citet{Duffy08} used the GADGET2 \citep{Springel05} code assuming the WMAP5 cosmology, fitting both NFW and Einasto profiles to the detected halos in the range $0 < z < 2$. They found that the concentration decreases with mass and redshift, but for high redshift ($z = 2$) this dependence on mass is less pronounced. When computing the relation for our sample we made use of the fitting formula for the NFW profile, full sample with $\Delta = 200$.

\citet{Prada12} has derived halo concentrations using the ratio between the maximum circular velocity to the virial velocity ($V_{max}/V_{200}$), modelling halos with a NFW profile. The study was based in the MultiDark, Bolshoi \citep{Klypin11}, Millennium I and II \citep{Springel05, Boylan09} simulations in a redshift range of $0 < z < 10$. The first two used WMAP5 and WMAP7 cosmological parameters while both Millenium simulations adopted WMAP1. They found an upturn in the c-M relation for increasing halo masses at high redshift.

Performing simulations with the PKDGRAV code \citep{Stadel01} and adopting the Planck cosmology, \citet{Dutton14} derived a c-M relation for halos in the range $0 < z < 5$. Density profiles were fitted to both NFW and Einasto models, with concentrations determined through these two profiles and also using the $V_{\rm{max}}/V_{200}$ ratio considering a NFW profile. They also have found an upturn at $z = 3$ when using $V_{\rm{max}}/V_{200}$ for measuring concentrations, but argued that this estimator is more sensitive to unrelaxed halos \citep{Ludlow12}, so care must be taken when interpreting the upturn feature. When using concentrations calculated from the NFW the derived relation is flat, and for the Einasto profile the slope of the relation is negative. Results for concentration are $20\%$ higher when adopting the Planck cosmology instead of WMAP1 values. We used the slope and zero point parameters derived for $M_{200}$ and $c_{200}$. 

\citet{Diemer15} studied the relation between concentration and peak height, the c-$\nu$ relation, for dark matter halos in $\Lambda$CDM and self-similar cosmologies. Most of the analysis was based in simulations consistent with WMAP7 cosmology. The concentrations were estimated by the \emph{ROCKSTAR} halo finder \citep{Behroozi13} which fits an NFW profile to mass distributions. They found an upturn at high $\nu$ argued to be associated with a higher fraction of unrelaxed halos, and a minimum concentration value that is dependent on redshift. We adopt the reported median of the best-fit parameters and the power spectrum from \emph{CAMB} to compute it and plotted the relation in terms of mass instead of peak height. 

\citet{Klypin16} used the suite of MultiDark simulations with different cosmological parameters to study dark matter halos. Fits to the density profiles were done using NFW and Einasto profiles with concentrations derived from the two profiles and from the ratio $V_{mas}/V_{200}$. They also have found an upturn in the c-M relation for massive halos at $z = 3$, speculating that this can be caused by the fact that the peaks in the Gaussian density field associated with the most massive clusters tend to be more spherical \citep{Doroshkevich70, Bardeen86}.

We derived the expected values of concentration predicted by each of the chosen numerical simulations at the mean redshift of the sample ($\bar{z} = 0.50$). Based on the results from simulations described in section \ref{simul_mc}, we found no evidence that the stacking procedure creates a non-negligible bias in the effective mass and concentration derived from our regression. We compared our constraints to a mean mass and concentration for a population of cluster following the same richness and redshift distributions as our sample. The result shows that measurement errors and statistical uncertainty are dominant. Therefore we assume that we can compare our stacked mass and concentration best-fit values and confidence interval to the theoretical predictions. 

\begin{figure}
\centering
 \includegraphics[width=8.8cm]{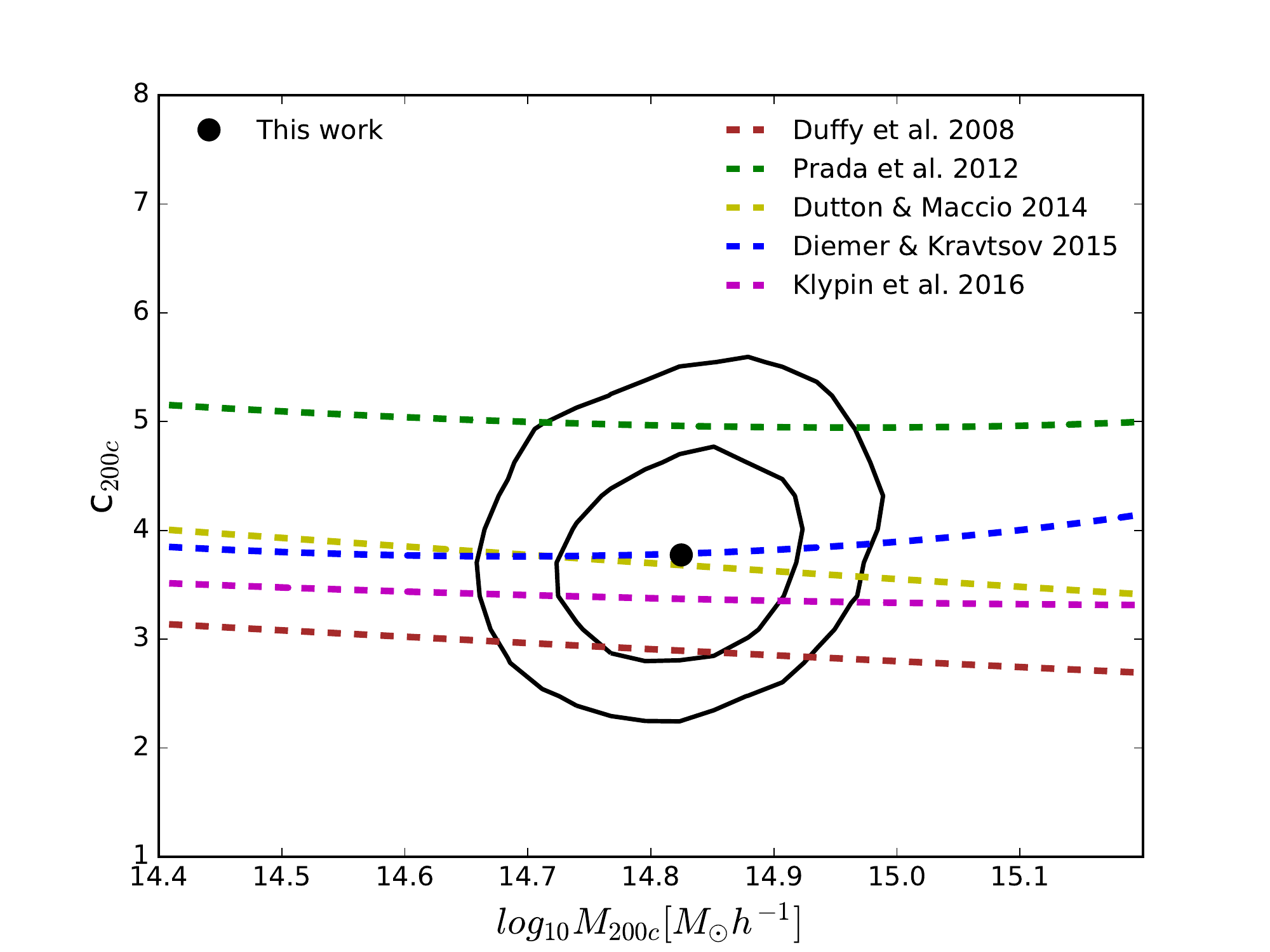}
 \caption{CODEX best-fit mass-concentration relation with the 1 and 2-sigma confidence levels. The lines represent the mean mass-concentration relation derived from different numerical predictions calculated at the mean redshift of our sample ($\bar{z} = 0.50$)}
\label{comp_simul}
\end{figure}

Figure \ref{comp_simul} shows that all relations can predict our observed concentration within $2\sigma$. The more recent relations from \citet{Dutton14}, \citet{Diemer15} and \citet{Klypin16} are in good agreement with our observational result; \citet{Duffy08} expectation value for $c_{200c}$ lies in the lower limit of our $1\sigma$ level, following the findings of previous studies (e.g \citealt{Covone14, Uitert16, Umetsu14, Okabe15} where $c_{200c}$ is under-predicted by the quoted relation; the prediction for $c_{200c}$ from \citet{Prada12} is higher for the given mass and redshift, being consistent with our constraint only at the $2\sigma$ level, and we note that our result falls in the beginning of the upturn region of their c-M relation for $z = 0.5$. 

\begin{figure}
\centering
 \includegraphics[width=8.8cm]{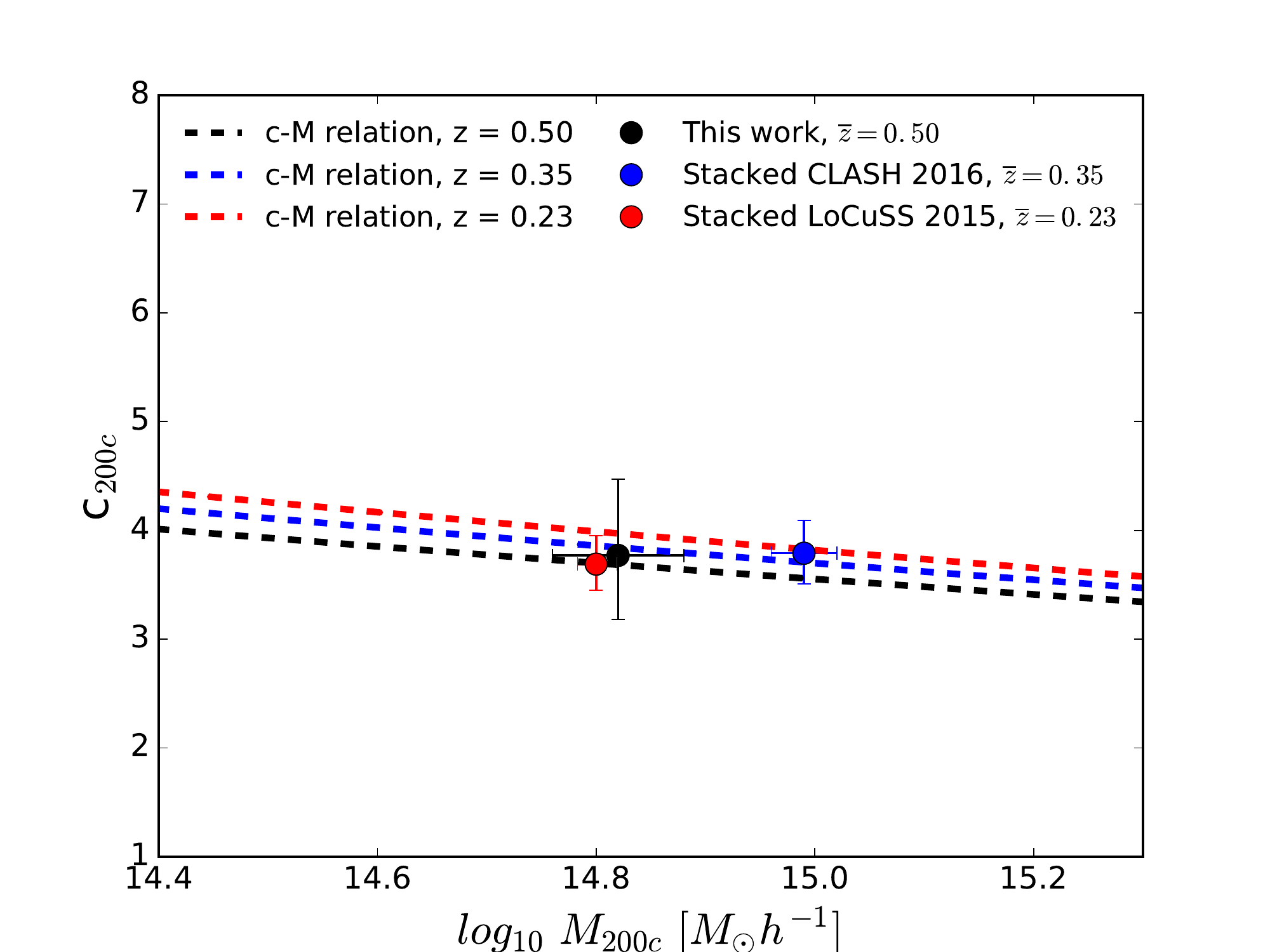}
 \caption{The concentration-mass relation predicted by \citet{Dutton14} evaluated at the mean redshift of our CFHT WL sample and at the mean redshift of the LoCuSS and CLASH stacked samples plotted with the best-fit c$_{200c}$ and M$_{200c}$ of each sample.}
\label{comp_simul2}
\end{figure}

We also analyse the expected c-M relation for different redshifts, evaluating the concentration predicted by \citet{Dutton14} using the mean redshift of our sample, the LoCuSS and the CLASH stacked samples. Results are shown in figure \ref{comp_simul2} together with the best-fit value of each study. The predicted evolution with redshift in the range spanned by the three studies is small and can not be confirmed by the quoted observational constraints. The concentration predicted for the LoCuSS sample is marginally consistent with the $1\sigma$ level constrained by the data, while the value predicted for the CLASH stacked sample is in good agreement with their observational result. 

\section{Summary and Conclusions}
\label{conclusion}

\ \ We have performed multi-band observations of 27 richness selected galaxy clusters at $0.40 \leqslant z \leqslant 0.62$ with the MegaCam instrument at CFHT. From this data we produced shear and photometric catalogues used in a weak lensing analysis. The study was based on the stacked shear profile built from the individual clusters and modelled by a three-component halo model including the off-centring effect and the two-halo term. This is a five parameters model with the free parameters $M_{200c}$ and $c_{200c}$ related to the main component, described by an NFW profile, the nuisance parameters $p_{\rm{cc}}$ (the fraction of clusters correctly centred) and $\sigma_{\rm{off}}$ (the off-set scale) determining the off-centring correction and the marginalised S$_m$ parameter to account for shape and distance measurement systematics. We select the background population and characterise the redshift distribution using the $p(z)$ derived from a colour-magnitude decision tree. We also applied a 5-band photometric redshift point estimator as a secondary method detailed in the appendix, used to cross check the robustness of the final results. The effective weighted number density of background sources is 6 gal/arcmin$^2$ and the significance of the lensing signal after stacking the individual clusters is $14\sigma$ computed using the full covariance matrix.

We constrain the free parameters through a Bayesian analysis with the parameters space explored by a MCMC algorithm. We let the two density profile parameters $M_{200c}$ and $c_{200c}$ be constrained by our data assuming uninformative priors and modelling the nuisance parameters related to the off-centring correction with priors based on X-ray and SZ data from \citet{Rykoff16}. The multiplicative parameter S$_m$ follows a Gaussian prior which assimilates the systematic uncertainties from shape and redshift measurements. The observed mass density profile is well described by the adopted halo model. 

The best-fit parameters for the stacked cluster sample based on the colour-magnitude decision tree are $M_{200c} = 6.6^{+1.0}_{-0.8}\ 10^{14} \ h^{-1} M_{\odot}$, $c_{200c} = 3.7^{+0.7}_{-0.6}$, quoted as the median of the posterior distribution with errors given by the interval containing $68 \%$ of the points. Results for the off-centring parameters are prior dominated, in particular for the fraction of clusters well-centred $p_{\rm{cc}}$ where the posterior distribution shows the same mean and width as the prior. The off-centre scale $\sigma_{\rm{off}}$ is affected by the data with the posterior presenting a broader distribution ($+ 12\%$, $- 30\%$) and mean value $20\%$ higher compared to the prior, although still compatible with the prior on the $1\sigma$ level.

 
We tested the impact on $M_{200}$ and $c_{200}$ resulting from different modellings of the off-centring parameters in the Bayesian analysis. The results given by the full model with off-centring priors based on observational data, when adopting flat priors and for fixed values of $p_{\rm{cc}}$ and $\sigma_{\rm{off}}$ are in agreement within the $1\sigma$ level. An uninformative prior for the off-centring parameters - specially for the parameter describing the fraction of clusters correctly centred $p_{\rm{cc}}$ - can lead to an overestimation of the off-centring effect and consequently to a bias toward higher values of the density profile parameters and a steeper concentration-mass relation, in agreement with what was argued in \citet{Du15}. To minimise this effect we consider a suitable approach to use the information about the goodness of the halo centred when available, as given by the redMapper parameter $P_{cen}$. We found that the redMaPPer expectation value for the fraction of well-centred halos for our sample is in agreement with the best-fit value of $p_{\rm{cc}}$ modelled by the adopted prior based on the observational results from \citet{Rykoff16}.

The results constrained by the two background selection methods are in good agreement (see Appendix \ref{ap:photoz} for the photometric redshift point estimator analysis). There is no evidence of systematic differences between the measurements based on the 5-band photometric redshift point estimator and on the p(z) distribution derived from the colour-magnitude decision tree, except potentially in the innermost part, where contamination by cluster member galaxies is significant. 
We consider the colour-magnitude decision tree to be a more robust method as it is based on deeper photometric information from the reference catalogue (including NIR), it uses the full p(z) information, it allows a more careful background selection and provides a better control over the systematic uncertainties. We note that the decision to apply a very conservative cut when selecting the background population, in order to avoid contamination by clusters members and foreground galaxies, leads to a lower background density compared to the photometric redshift point estimator cut. The good agreement between the methods however indicates that the photometric redshift approach appears to lead to no significant contamination by foreground or cluster member galaxies and that the final bias due to scatter in the photo-z estimates plays a smaller role than what would be expected from the worst case simulations.

When comparing our best-fit values for $M_{200}$ and $c_{200}$ with the stacked results presented in \citet{Okabe15} (LoCuSS) and \citet{Clash16} (CLASH) we found no evidence for a evolution in the concentration-mass relation between the mean redshift of the samples, $\bar{z} = 0.50$, $\bar{z} = 0.35$ and $\bar{z} = 0.23$ respectively. Differences in the halo model adopted for each analysis could partially explain this fact as well as selection effects, which could affect especially the CLASH sample. We note, however, that the expected evolution for the concentration between the given redshifts and X-ray related selection biases on $c$ are small and falls within our $1\sigma$ confidence level. 

\citet{Clash16} (CLASH) found a higher value of mass for the stacked sample, compatible with the average higher X-ray luminosity of their systems. The mass constrained for the stacked sample in \citet{Okabe15} (LoCuSS) is comparable to our result, where we have obtained marginally larger values of mass. This result is surprising given the expected evolution of the L$_X$-M relation. The similarity of the mass range and mean L$_X$ between the CODEX and LoCuSS samples suggests either that the evolution of the Lx-M relation is shallower than self-similar, or that there is an evolution in the scatter \citep{Mantz16a}. Detailed X-ray studies would help to investigate the origin of the apparent lack of a strong evolution in L$_X$ for the mass range probed by CODEX and LoCuSS.

A comparison between the luminosity distributions in figures \ref{lx_full} and \ref{codexmEz} shows that by applying a L$_X$ cut, as done by LoCuSS, one samples only a small part of the full L$_X$ distribution given the richness range, corresponding to the highest-L$_X$ systems. Also from the full CODEX luminosity distribution (figure \ref{lx_full}) we see that the average and the scatter on L$_X$ are not constant with redshift, with lower z systems spanning a larger range of L$_X$ and having on average lower value of L$_X$. \citealt{Mantz16a} discussed the evolution of the scatter in X-ray luminosity pointing to the late assembly of dense, bright cores as an explanation for the decreasing scatter with z.


Numerical predictions based on N-body simulations for the concentration-mass relation are in agreement with our observational result within the $2\sigma$ confidence level. For each numerical simulations we calculate the expected concentration given a mass range at the mean redshift of our sample. The expected value of concentration given by \citet{Duffy08} is marginally consistent with our $1\sigma$ level, tending to predict a lower value of concentration. The concentration derived from \citet{Prada12} is too high and consistent with our observational result only at the $2\sigma$ level. The more recent relations derived in \citet{Dutton14} and \citet{Klypin16} and the relation between concentration and peak height (c-$\nu$) presented in \citet{Diemer15} are in good agreement with our best-fit values. We conclude that the relations derived in \citet{Dutton14}, \citet{Diemer15} and \citet{Klypin16} are a good description of our data and can be used to predict concentration values for individual CODEX weak lensing mass measurements, which in turn will be used to derive the CODEX scaling-relations. 


\section*{Acknowledgements}
 
\ \ This work is based on observations obtained with MegaPrime/MegaCam, a joint project of CFHT and CEA/IRFU, at the Canada-France-Hawaii Telescope (CFHT) which is operated by the National Research Council (NRC) of Canada, the Institut National des Science de l'Univers of the Centre National de la Recherche Scientifique (CNRS) of France, and the University of Hawaii.

NC acknowledges financial support from the Brazilian agencies CNPQ and CAPES (process \#2684/2015-2 PDSE). NC also acknowledges support from the Max-Planck-Institute for Extraterrestrial Physics. This research was supported by the DFG cluster of excellence "Origin and Structure of the Universe" (www.universe-cluster.de). ESC acknowledges financial support from Brazilian agencies CNPQ and FAPESP (process \#2014/13723-3). Support for DG was provided by NASA through the Einstein Fellowship program, grant PF 5-160138. LM acknowledges STFC grant ST/N000919/1. AF \& CK acknowledge the Finnish Academy award, decision 266918. RAD thanks support from CAPES grant PPVE 23038.008197/2012-45 and CNPq grant 312307/2015-2.

We would like to thank C. Heymans and Noburo Okabe for useful discussions. We acknowledge R. Bender for the use of his photometric redshift pipeline in this work. We thank Matthew R. Becker and Andrej Kravtsov for making their cluster simulations available. NC acknowledge J. Weller for the hospitality.

\appendix

\section{Photometric redshift point estimator}
\label{ap:photoz}

\ \ In this section we perform the WL analysis based on the photometric redshift point estimator. We follow the same steps described in section \ref{wl}, with the difference that we do not include the parameter $S_m$ as we are not able to quantify the systematic uncertainties from this method. Although the photo-z point estimator cut is considered to be more sensitive to systematics due to the scatter in estimates, being affected by foreground and cluster members contamination, we find a good agreement with the results from the conservative colour-magnitude decision tree method. 

A direct evaluation of the photometric redshift accuracy is not possible due to lack of a representative spectroscopic sample (the SDSS spectroscopic sample is too special for a representative comparison). We therefore estimate our redshift errors based on the results of \citet{Brimioulle13}, given their analysis considered a similar field (CFHTLS wide) observed with same instrument and filter set and using the same template fitting code (Bender et al. 2001) for the photometric redshift estimates. They found for the faintest galaxies (22.5 $\leqslant$ mag${_i'}$ $\leqslant$ 24) an outlier rate of $\eta \sim 5\%$ and a photometric redshift scatter of $\Delta_{z/(1+z)} \sim 0.05$.

To select the background sources based on the 5-bands CFHT photometric redshift point estimator we apply a cut following the prescription detailed in \citet{Brimioulle13}:

\[
1.1 z_{\rm{cluster}} + 0.15 < z_{\rm{gal}} < 2,
\]

\noindent where $z_{\rm{cluster}}$ is the spectroscopic redshift of each  cluster and $z_{\rm{gal}}$ is the photometric redshift assigned to each galaxy in the field. This criterion is motivated by the results from simulations performed in \citet{Brimioulle13}: when converting the shear profile to a mass density profile, the fractional error depends on the ratio between the error and the value of the critical density (eq. \ref{scrit}). Adopting Gaussian errors of $0.05(1+z)$ for the photometric redshift in the simulations, they found that this ratio is below 0.3 if $z_{\rm{gal}} > 1.1 z_{\rm{cluster}} + 0.15$ and $z_{\rm{cluster}} > 0.05$. Applying this criterion leads to a background population with $\beta > 0.1$; by excluding background galaxies close to the cluster redshift (and therefore $\beta < 0.1$) we avoid the low signal-to-noise shear signal which can also be contaminated by cluster member galaxies. The higher cut is applied to eliminate less reliable photometric redshift estimate. For $z_{\rm{gal}} > 2$ we may select a population containing foreground objects scattered to the background. In addition we expect, due to this cut, a systematic error of less then 2 per cent which we confirm by simulations, as discussed in Section \ref{errorphotoz}.

\subsection{Photometric redshift systematics}
\label{errorphotoz}

\ \ Photometric redshift errors can affect the result of weak lensing analyses. Already symmetric photo-z errors  without bias can impose a bias on the measured weak lensing mass. We therefore create simulated weak lensing catalogues in order to estimate to which extent the result might be affected. For this we use the known cluster and background positions and assume the measured photometric redshift to be true. This enables us to calculate the expected analytical weak lensing shear for our background objects. In the next step we add a symmetric photometric redshift scatter to the original values to simulate photo-z errors and reanalyse the obtained simulated sample. We used several modellings for this photo-z scatter (constant scatter for all objects of $0.03(1+z)$ or $0.05(1+z)$, respectively, but also a progressive scatter function being $0.03(1+z)$ for objects with $i \le 22.5$ and increasing to $0.1(1+z)$, taking into account the photometric redshift error typically increases with decreasing signal-to-noise).

Depending on the modelling we observe mass biases from around 5 per cent up to values of 20 per cent in the worst case. However, it is not trivial to obtain the correct modelling function, since the redshift uncertainties are simultaneously functions of galaxy magnitude, redshift and spectral type. So in general the redshift accuracy deteriorates for fainter, bluer and more distant galaxies.

\subsection{Results}
\label{ap:resul}

\ \ We now present the results for the Bayesian analysis based on the photometric redshift point estimator method. We follow the same approach described in section \ref{wl}, first stacking the 27 clusters and then modelling the shear profile with three components, using the same covariance matrix and profile fitting. The only difference is the absence of the multiplicative parameter $S_m$ now, as we are not able to characterise the systematic uncertainty from the photo-z point estimator selection. 

\begin{figure}
\centering
\includegraphics[width=8cm]{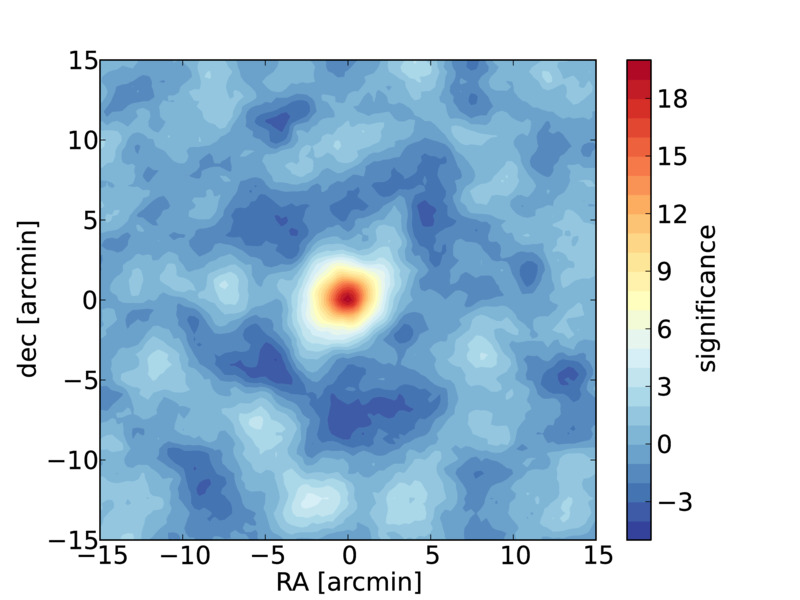} 
 \caption{Aperture mass significance map for the stacked cluster derived for the point estimator photometric redshift method.}
\label{fig:massmapphotoz}
\end{figure}

The aperture map for the stacked sample, calculated as described in section \ref{stack}, is shown in figure \ref{fig:massmapphotoz}. The significance peak of the aperture mass is $20 \sigma$. Using the full covariance matrix we detect the signal with a S/N of $13\sigma$ (eq. \ref{sn_matrix}).

\begin{figure}
\centering
\includegraphics[width=8.8cm]{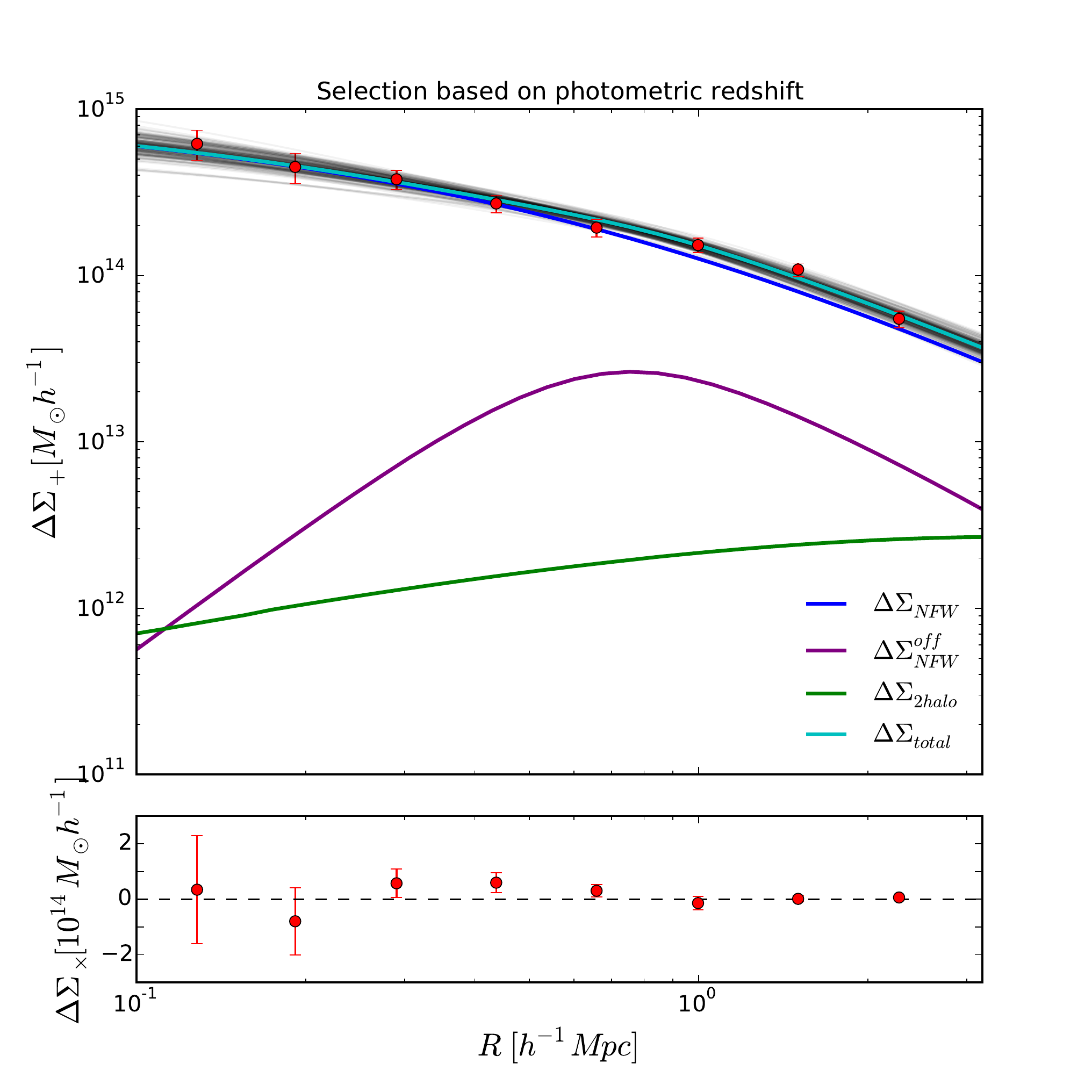} 
 \caption{Stacked tangential shear profile of the 27 clusters based on the photo-z point estimator method, in units of projected mass density, with 100 randomly selected MCMC realisations to sample the $\Delta \Sigma_{total}$ variance (grey lines). Each component of the fitted halo model, as well as the total signal, is plotted separately. The lower panel shows the cross component of the shear signal, used to check for systematics.}
\label{full_photoz}
\end{figure}

Figure \ref{full_photoz} shows the observed density profile (red dots) with errors given by the square root of the main diagonal of the covariance matrix and each individual component as well as the final profile given by equation \ref{h_model}. The grey lines show 100 random MCMC realisations exemplifying the parameter space explored by the fitting.

\begin{figure}
\centering
\includegraphics[width=8.2cm]{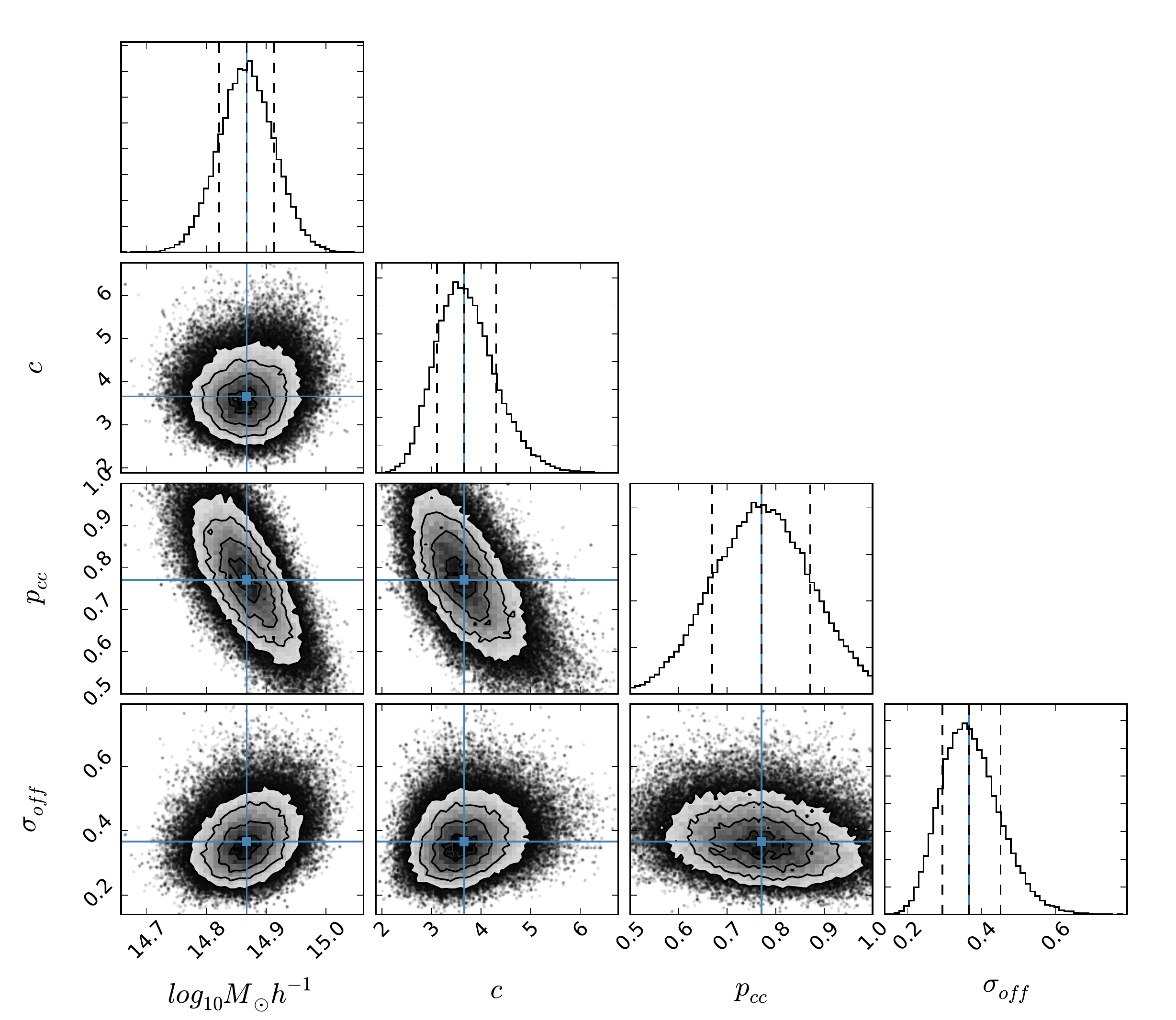} 
 \caption{One and two dimensional projections of the posterior distributions for the stacked profile parameters M$_{200c}$, c$_{200c}$ and the off-set parameters based on the photo-z point estimator. The blue dots represent the best-fit values quoted as the median of the distributions.}
\label{triangle_photoz}
\end{figure}

In table \ref{tab:fitphotoz} we quote the median and the interval containing 68$\%$ of the points from the posterior probability distribution of the MCMC chains. We also present the mode of the distributions and the values given by a Gaussian approximation. Figure \ref{triangle_photoz} shows the 1 and 2D posterior distributions of the free parameters.

\begin{table*}
\caption{Summary of the posterior distributions for the photo-z point estimator. Should be compared to Table \ref{tab:fit}.}
\begin{tabular}{|p{4.5cm}| p{2cm} p{2cm} p{2cm} p{2cm}|}
\hline
Adopted value & M$_{200c}$ $[10^{14} h^{-1} M_{\odot}]$ & c$_{200c}$ & $p_{\rm{cc}}$ & $\sigma_{\rm{off}}$ $[h^{-1} Mpc]$ \\
\hline
Median and 68\% interval & $6.9^{+1.0}_{-0.9}$ & $3.5^{+0.6}_{-0.5}$ & $0.8^{+0.1}_{-0.1}$ & $0.36^{+0.08}_{-0.07}$ \\
Mode of the 1D distribution & $6.6$ & $3.6$ & $0.8$ & $0.42$ \\
Gaussian approximation & $6.9^{+1.0}_{-0.9}$ & $3.6 \pm 0.6$ & $0.8 \pm 0.1$ & $0.36 \pm 0.08$ \\
\hline
\end{tabular}
\label{tab:fitphotoz}
\end{table*}

\begin{figure}
\centering
\includegraphics[width=8.2cm]{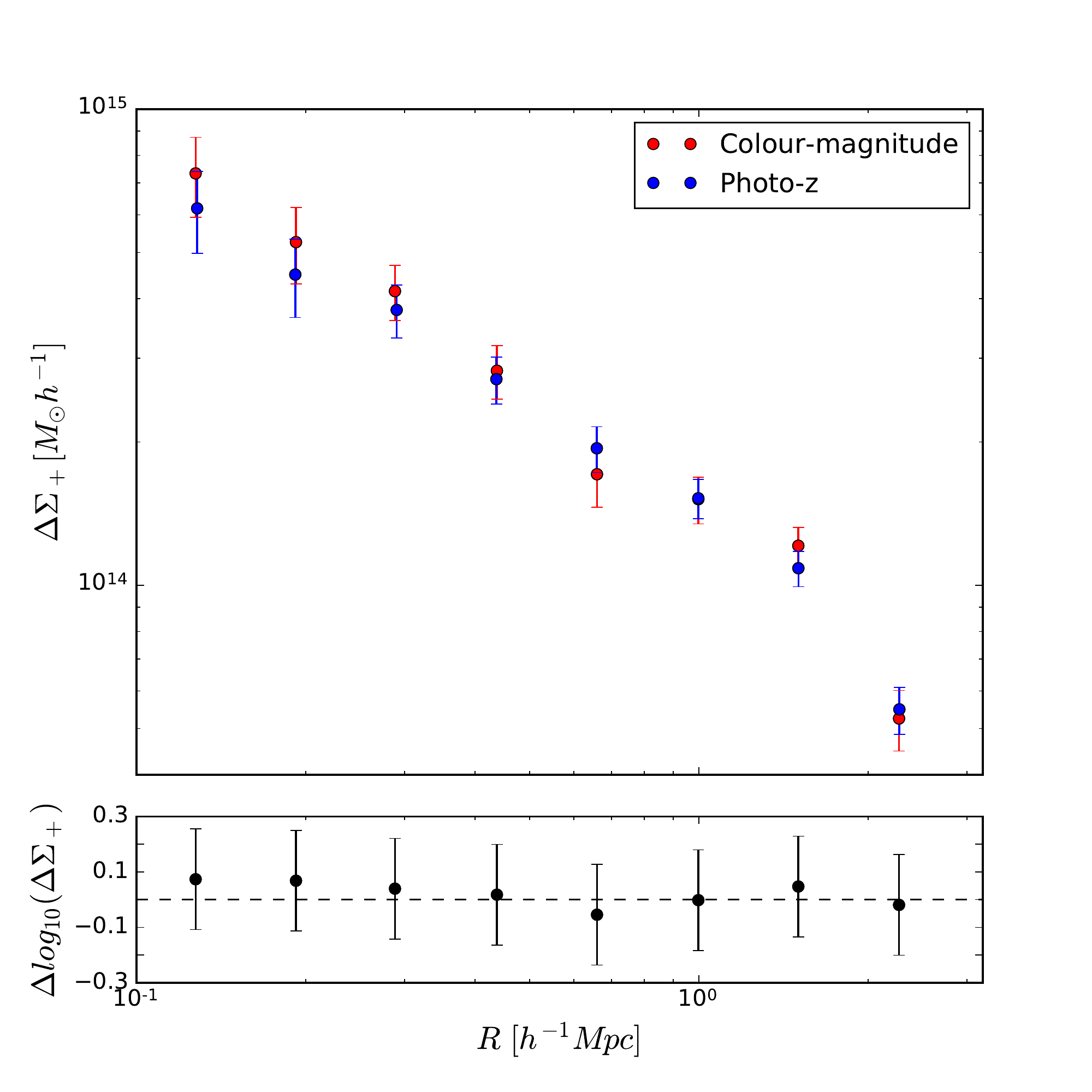} 
 \caption{The stacked tangential shear profile obtained by the two different background selection methods. Errors are given by the square root of the main diagonal of the covariance matrix derived for each method. The lower panel shows the residual between the two profiles.}
\label{profs}
\end{figure}

This result is in good agreement with the one constrained by the colour-magnitude decision tree (should be compared with table \ref{tab:fit}), with no evidence of a pronounced bias caused by the point estimator cut. The density profiles derived from each method are consistent within the uncertainties, as can be seen in figure \ref{profs}. The photo-z point estimator method shows a tendency to be smaller on small scales ($<500$ kpc), where one would expect the leakage of cluster galaxies into the background sample to be most severe. The conservative cut applied to the colour-magnitude method is more robust in excluding these cluster galaxies.

\section{Background population properties}

\ \ We present the properties of the background galaxies selected by the two methods, the colour-magnitude decision tree and the photometric redshift point estimator, for each cluster in Table \ref{tab:comp}. Note that the colour-magnitude decision tree gives higher values of the lensing quantity $\beta$ and a median redshift of the sources that is also higher compared to the photo-z method, whereas the number (N$_{\rm{bg}}$) and the weighted number density (N$_{\rm{eff}}$) of background galaxies is lower. These numbers reflect the very conservative criteria adopted for the colour-magnitude decision tree method in order to better control the systematic uncertainties arising from contamination of the source sample by cluster member and foreground galaxies.  


\begin{table*}
\caption{Background properties of the CODEX stacked weak lensing clusters sample.}
\label{tab:comp}
\begin{tabular}{l l | c c c c | c c c c}
\hline
 &  & \multicolumn{4}{c}{magnitude decision tree} & \multicolumn{4}{c}{Photo-z point estimate} \\
\hline
ID & z$_{cl}$ & N$_{bg}$ & N$_{eff}$ & $\widetilde{z_{bg}}$ & $\overline{\beta_{eff}}$ & N$_{bg}$ & N$_eff$ & $\widetilde{z_{bg}}$ & $\overline{\beta_{eff}}$ \\

\hline
12451 & 0.58 & 14516 & 4.86 & 1.13 & 0.41 & 22407 & 7.62 & 1.02 & 0.36 \\
24865 & 0.49 & 18228 & 5.62 & 1.02 & 0.45 & 25339 & 7.88 & 0.88 & 0.40 \\
24872 & 0.40 & 14924 & 5.27 & 0.96 & 0.51 & 25832 & 9.10 & 0.86 & 0.46 \\
24877 & 0.59 &  9795 & 3.27 & 1.13 & 0.40 & 18017 & 6.25 & 0.95 & 0.33 \\
24981 & 0.51 & 16197 & 4.92 & 1.04 & 0.52 & 26000 & 7.98 & 0.89 & 0.48 \\
25424 & 0.51 & 27563 & 8.52 & 1.11 & 0.45 & 39530 & 12.27 & 0.96 & 0.42 \\
25953 & 0.48 & 24892 & 6.81 & 1.04 & 0.46 & 37064 & 10.21 & 0.9 & 0.42 \\
27940 & 0.45 & 16673 & 5.20 & 0.96 & 0.46 & 22644 & 7.14 & 0.87 & 0.42 \\ 
27974 & 0.47 & 20410 & 7.73 & 1.04 & 0.46 & 28729 & 10.91 & 0.89&  0.42 \\
29283 & 0.55 & 15434 & 5.63 & 1.09 & 0.43 & 24211 & 8.92 & 0.96 & 0.37 \\
29284 & 0.55 & 15344 & 5.60 & 1.10 & 0.43 & 22420 & 8.26 & 0.98 & 0.38 \\
29811 & 0.49 & 25856 & 7.91 & 1.10 & 0.46 & 34838 & 10.72 & 0.92 & 0.42 \\
35361 & 0.41 & 27783 & 7.37 & 1.01 & 0.51 & 49141 & 13.05 & 0.88 & 0.47 \\   
35399 & 0.52 & 19317 & 5.64 & 1.10 & 0.44 & 29637 & 8.69 & 0.94 & 0.40 \\
36818 & 0.58 & 12931 & 5.78 & 1.14 & 0.42 & 18080 & 8.17 & 1.01 & 0.37 \\
37098 & 0.54 & 10114 & 3.48 & 1.11 & 0.43 & 14501 & 5.07 & 0.95 & 0.38 \\
41843 & 0.43 & 19126 & 5.70 & 0.97 & 0.49 & 29030 & 8.71 & 0.87 & 0.44 \\
43403 & 0.42 & 23692 & 7.85 & 1.02 & 0.50 & 34369 & 11.52 & 0.87 & 0.46 \\
46649 & 0.62 & 13956 & 4.25 & 1.19 & 0.40 & 30498 & 9.53 & 0.95 & 0.30 \\
47981 & 0.54 & 10846 & 4.29 & 1.13 & 0.44 & 14002 & 5.66 & 0.95 & 0.37 \\
50492 & 0.53 & 15195 & 4.69 & 1.11 & 0.43 & 24163 & 8.70 & 0.87 & 0.41 \\
50514 & 0.47 & 18150 & 6.41 & 0.99 & 0.45 & 22437 & 7.03 & 0.95 & 0.38 \\
52480 & 0.56 & 10693 & 3.26 & 1.10 & 0.41 & 17670 & 5.53 & 0.94 & 0.35 \\ 
54795 & 0.43 & 19774 & 5.98 & 0.98 & 0.49 & 27909 & 8.54 & 0.87 & 0.45 \\
55181 & 0.55 & 16741 & 4.58 & 1.11 & 0.44 & 24069 & 6.75 & 0.97 & 0.38 \\
59915 & 0.47 & 25855 & 6.88 & 1.03 & 0.46 & 36966 & 9.90 & 0.89 & 0.42 \\
64232 & 0.53 & 12477 & 2.97 & 1.10 & 0.43 & 22906 & 5.56 & 0.9 & 0.37 \\
\hline 
\end{tabular}
\end{table*}

\clearpage

\onecolumn
\section{Observational information}

\begin{longtable}{@{}cccccc@{}}
\caption[]{Observational properties of the CODEX stacked weak lensing sample. The fields 24877 and 29284 have a secondary cluster, 24872 and 29283 respectively.} \\
\label{tab:obs} \\

\hline 
ID & Filter & Seeing & N. of exp. & Exp. time & $m_{lim}$ \tabularnewline \hline
    \endfirsthead
    
\multicolumn{6}{c}%
    {{\bfseries \tablename\ \thetable{} -- continued from previous page}} \\
    \hline
    \endhead
   
    \hline
    \multicolumn{6}{c}{{Continues on next page}} \tabularnewline \hline
    \endfoot
    
    \hline
    \endlastfoot

\hline
12451 & g.MP9401 & 1.08 & 05 & 1050.91 & 25.12 \\
12451 & i.MP9702 & 0.61 & 15 & 8403.23 & 24.89 \\ 
12451 & r.MP9601 & 0.94 & 09 & 2491.80 & 25.01 \\
12451 & u.MP9301 & 0.83 & 05 & 2601.08 & 25.20 \\
12451 & z.MP9801 & 1.04 & 02 & 1080.51 & 22.56 \\
24865 & g.MP9401 & 0.91 & 06 & 1261.01 & 25.16 \\
24865 & i.MP9702 & 0.73 & 07 & 3921.39 & 24.71 \\
24865 & r.MP9601 & 0.75 & 05 & 1350.90 & 24.63 \\
24865 & u.MP9301 & 0.79 & 05 & 2600.98 & 25.24 \\
24865 & z.MP9801 & 0.68 & 03 & 1620.65 & 23.12 \\
24877 & g.MP9401 & 0.97 & 03 & 1020.58 & 25.13 \\
24877 & i.MP9702 & 0.84 & 08 & 4481.69 & 24.77 \\
24877 & r.MP9601 & 0.87 & 03 & 1494.46 & 24.81 \\
24877 & u.MP9301 & 1.05 & 05 & 2800.96 & 25.32 \\
24877 & z.MP9801 & 0.88 & 04 & 2240.88 & 23.59 \\
24981 & g.MP9401 & 0.65 & 05 & 1600.93 & 25.32 \\
24981 & i.MP9701 & 0.89 & 04 &  960.73 & 23.80 \\
24981 & i.MP9702 & 0.60 & 06 & 3361.26 & 24.59 \\
24981 & r.MP9601 & 0.64 & 12 & 7406.18 & 25.49 \\
24981 & u.MP9301 & 0.93 & 05 & 2600.99 & 25.30 \\
24981 & z.MP9801 & 0.81 & 04 & 1440.78 & 22.91 \\
25424 & g.MP9401 & 0.76 & 04 &  840.56 & 24.99 \\
25424 & i.MP9702 & 0.65 & 08 & 4481.58 & 24.73 \\
25424 & r.MP9601 & 0.65 & 05 & 1350.77 & 24.51 \\
25424 & u.MP9301 & 1.03 & 15 & 7802.94 & 25.73 \\
25424 & z.MP9801 & 0.46 & 04 & 2160.71 & 23.27 \\
25953 & g.MP9401 & 0.91 & 05 & 1050.84 & 25.21 \\
25953 & i.MP9702 & 0.72 & 08 & 4481.55 & 24.92 \\
25953 & r.MP9601 & 0.84 & 04 &  990.68 & 24.58 \\
25953 & u.MP9301 & 0.85 & 04 & 2080.73 & 25.09 \\
25953 & z.MP9801 & 0.83 & 04 & 2160.81 & 23.62 \\
27940 & g.MP9401 & 0.68 & 03 & 1080.49 & 25.04 \\
27940 & i.MP9702 & 0.91 & 09 & 5041.67 & 24.77 \\
27940 & r.MP9601 & 0.57 & 03 & 1500.49 & 24.33 \\
27940 & u.MP9301 & 0.81 & 05 & 2860.72 & 25.21 \\ 
27940 & z.MP9801 & 0.86 & 04 & 2372.79 & 23.36 \\
27974 & g.MP9401 & 0.65 & 03 & 1080.46 & 25.01 \\
27974 & i.MP9702 & 0.44 & 08 & 4481.20 & 24.65 \\
27974 & r.MP9601 & 0.67 & 03 & 1500.44 & 24.31 \\
27974 & u.MP9301 & 0.82 & 04 & 2288.71 & 25.23 \\
27974 & z.MP9801 & 0.86 & 04 & 2372.81 & 23.29 \\
29284 & g.MP9401 & 1.20 & 06 & 1260.90 & 25.23 \\
29284 & i.MP9702 & 0.56 & 08 & 4481.56 & 24.84 \\
29284 & r.MP9601 & 0.76 & 03 &  930.43 & 24.45 \\
29284 & u.MP9301 & 0.90 & 05 & 2600.91 & 25.22 \\
29284 & z.MP9801 & 0.87 & 04 & 2160.60 & 23.24 \\
29811 & i.MP9702 & 0.44 & 06 & 3361.00 & 24.51 \\
29811 & r.MP9601 & 0.78 & 08 & 1981.05 & 24.86 \\
29811 & u.MP9301 & 0.84 & 04 & 2080.59 & 24.85 \\
29811 & z.MP9801 & 0.51 & 04 & 2160.57 & 23.26 \\
35361 & g.MP9401 & 0.96 & 05 & 1050.67 & 25.06 \\
35361 & i.MP9702 & 0.57 & 07 & 3921.26 & 24.68 \\
35361 & r.MP9601 & 0.94 & 06 & 1560.88 & 24.63 \\
35361 & u.MP9301 & 0.77 & 10 & 5202.05 & 25.49 \\
35361 & z.MP9801 & 0.83 & 04 & 2320.82 & 23.14 \\
35399 & g.MP9401 & 0.62 & 03 &  900.69 & 24.81 \\
35399 & i.MP9702 & 0.55 & 08 & 4481.22 & 24.58 \\
35399 & r.MP9601 & 0.67 & 03 & 1380.42 & 24.39 \\
35399 & u.MP9301 & 0.94 & 07 & 3641.14 & 24.94 \\
35399 & z.MP9801 & 0.88 & 04 & 2160.70 & 23.35 \\
36818 & g.MP9401 & 0.82 & 03 & 1020.47 & 25.03 \\
36818 & i.MP9702 & 0.45 & 10 & 5601.63 & 24.73 \\
36818 & r.MP9601 & 0.88 & 03 & 1500.50 & 24.48 \\
36818 & u.MP9301 & 1.07 & 05 & 2860.87 & 25.29 \\
36818 & z.MP9801 & 0.63 & 10 & 5883.62 & 23.69 \\
37098 & g.MP9401 & 0.65 & 03 & 1020.46 & 25.10 \\
37098 & i.MP9702 & 0.70 & 08 & 4481.28 & 24.66 \\
37098 & r.MP9601 & 0.60 & 03 & 1500.41 & 24.69 \\
37098 & u.MP9301 & 0.85 & 05 & 2860.86 & 25.25 \\
37098 & z.MP9801 & 0.49 & 04 & 2372.67 & 23.38 \\
41843 & g.MP9401 & 0.71 & 03 & 1020.41 & 25.12 \\
41843 & i.MP9702 & 0.73 & 08 & 4481.31 & 24.90 \\
41843 & r.MP9601 & 0.67 & 06 & 2988.84 & 24.98 \\
41843 & u.MP9301 & 0.83 & 05 & 2800.70 & 25.30 \\
41843 & z.MP9801 & 0.67 & 08 & 4481.21 & 23.52 \\
43403 & g.MP9401 & 0.87 & 05 & 1070.61 & 25.06 \\
43403 & i.MP9702 & 0.54 & 08 & 4481.31 & 24.70 \\
43403 & r.MP9601 & 0.87 & 03 &  630.37 & 24.25 \\
43403 & u.MP9301 & 0.84 & 05 & 2600.61 & 25.22 \\
43403 & z.MP9801 & 0.41 & 02 & 1080.21 & 22.47 \\
46649 & g.MP9401 & 0.62 & 03 & 1020.34 & 24.76 \\
46649 & i.MP9702 & 0.51 & 09 & 5041.11 & 24.67 \\
46649 & r.MP9601 & 0.82 & 04 & 1992.43 & 24.68 \\
46649 & u.MP9301 & 0.79 & 05 & 2800.74 & 24.96 \\ 
46649 & z.MP9801 & 0.77 & 04 & 2240.51 & 23.44 \\
47981 & g.MP9401 & 1.17 & 03 & 1020.22 & 25.08 \\
47981 & i.MP9702 & 0.73 & 09 & 5041.09 & 24.81 \\
47981 & r.MP9601 & 0.65 & 04 & 1992.33 & 24.86 \\
47981 & u.MP9301 & 1.21 & 05 & 2800.48 & 25.20 \\
47981 & z.MP9801 & 0.53 & 04 & 2240.44 & 23.27 \\
50492 & g.MP9401 & 0.56 & 03 & 1080.27 & 24.91 \\
50492 & i.MP9702 & 0.70 & 08 & 4481.05 & 24.58 \\
50492 & r.MP9601 & 0.42 & 06 & 3000.85 & 24.87 \\
50492 & u.MP9301 & 0.56 & 05 & 2800.69 & 25.16 \\
50492 & z.MP9801 & 0.58 & 04 & 2240.54 & 23.11 \\
50514 & g.MP9401 & 0.78 & 03 & 1020.35 & 25.04 \\
50514 & i.MP9702 & 0.72 & 08 & 4481.18 & 24.80 \\
50514 & r.MP9601 & 0.71 & 05 & 2490.61 & 24.67 \\
50514 & u.MP9301 & 0.95 & 05 & 2800.71 & 25.24 \\
50514 & z.MP9801 & 0.59 & 04 & 2240.55 & 23.20 \\
52480 & g.MP9401 & 0.79 & 03 &  900.37 & 25.00 \\
52480 & i.MP9702 & 0.76 & 10 & 5600.96 & 24.71 \\
52480 & r.MP9601 & 0.94 & 03 & 1380.30 & 24.56 \\
52480 & u.MP9301 & 0.71 & 05 & 2600.46 & 25.18 \\
52480 & z.MP9901 & 0.81 & 04 & 2240.62 & 22.91 \\
54795 & g.MP9401 & 1.13 & 03 & 1020.32 & 24.97 \\
54795 & i.MP9702 & 0.64 & 08 & 4480.77 & 24.55 \\
54795 & r.MP9601 & 0.74 & 03 & 1494.36 & 24.16 \\
54795 & u.MP9301 & 0.94 & 05 & 2800.52 & 25.13 \\
54795 & z.MP9801 & 0.70 & 04 & 2240.48 & 23.18 \\
55181 & g.MP9401 & 1.19 & 03 & 1080.28 & 24.94 \\
55181 & i.MP9702 & 0.62 & 08 & 4480.72 & 24.56 \\
55181 & r.MP9601 & 0.74 & 06 & 3000.57 & 24.75 \\
55181 & u.MP9301 & 1.18 & 05 & 2800.71 & 25.02 \\
55181 & z.MP9801 & 0.67 & 04 & 2240.35 & 22.99 \\
59915 & i.MP9702 & 0.62 & 08 & 4480.72 & 24.80 \\
59915 & r.MP9601 & 0.91 & 03 & 1494.27 & 24.50 \\
59915 & u.MP9301 & 0.96 & 05 & 2800.45 & 25.08 \\
59915 & z.MP9801 & 0.76 & 04 & 2240.36 & 23.33 \\
64232 & g.MP9401 & 0.65 & 03 & 1020.23 & 25.05 \\
64232 & i.MP9702 & 0.79 & 06 & 3480.57 & 24.60 \\
64232 & r.MP9601 & 0.84 & 03 & 1494.28 & 24.59 \\
64232 & u.MP9301 & 0.90 & 05 & 2800.49 & 25.20 \\
64232 & z.MP9801 & 0.74 & 04 & 2240.38 & 23.17 \\
\hline 

\end{longtable}



%
%




\label{lastpage}

\end{document}